\documentclass{article}
\usepackage{PRIMEarxiv}
\usepackage{graphicx} 
\usepackage[sort&compress]{natbib}
\usepackage[utf8]{inputenc} 
\usepackage[T1]{fontenc}    
\usepackage[colorlinks = True, linkcolor = red, citecolor = violet]{hyperref}       
\usepackage{booktabs}       
\usepackage{amsfonts}       
\usepackage{nicefrac}       
\usepackage{microtype}      

\usepackage{stmaryrd}
\usepackage{amssymb}
\usepackage{amsmath}
\usepackage{amsthm}
\usepackage{graphicx} 
\usepackage{placeins} 
\usepackage{algorithmic}
\usepackage{float}
\usepackage{pifont}
\usepackage{cleveref}
\usepackage{bm}
\usepackage{mathtools}
\usepackage[ruled,vlined]{algorithm2e} 
\usepackage{wrapfig}
\usepackage{stmaryrd}
\usepackage[HTML]{xcolor}
\usepackage[most]{tcolorbox}

\definecolor{myboxframe_bio}{HTML}{D16F7D}  
\definecolor{myboxback_bio}{HTML}{FDEEEF}   
\definecolor{myboxframe_methodo}{HTML}{EA5B0C}  
\definecolor{myboxback_methodo}{HTML}{FFF0E6}   

\newcommand{\x}{\bm{x}}
\newcommand{\R}{\mathbb{R}}
\newcommand{\y}{\bm{y}}
\newcommand{\Y}{\mathbf{Y}}
\newcommand{\ff}{\bm{f}}
\newcommand{\bTheta}{\boldsymbol{\Theta}}
\newcommand{\bXi}{\boldsymbol{\Xi}}

\DeclarePairedDelimiterX{\norm}[1]{\lVert}{\rVert}{#1}
\newtheorem{definition}{Definition}

\title{
Data-Driven Discovery of Digital Twins in Biomedical Research
}

\author{
  Clémence Métayer \\
  Inserm U1331, Institut Curie \\
  Saint-Cloud, France \\
  \texttt{clemence.metayer@curie.fr} \\
  \And
  Annabelle Ballesta\thanks{These authors equally supervised this work.} \\
  Inserm U1331, Institut Curie \\
  Saint-Cloud, France \\
  \AND
  Julien Martinelli\footnotemark[1]\\
  Aalto University\\
  Espoo, Finland
}
\date{}

\begin{document}
\maketitle
\begin{abstract}
Recent technological advances have expanded the availability of high-throughput biological datasets, enabling the reliable design of digital twins of biomedical systems or patients. Such computational tools represent key chemical reaction networks driving perturbation or drug response and can profoundly guide drug discovery and personalized therapeutics. Yet, their development still depends on laborious data integration by the human modeler, so that automated approaches are critically needed. 
The successes of data-driven system discovery in Physics, rooted in clean datasets and well-defined governing laws, have fueled interest in applying similar techniques in Biology, which presents unique challenges. Here, we reviewed existing methodologies for automatically inferring digital twins from biological time series, which mostly involved symbolic or sparse regression. We evaluated algorithms according to eight biological and methodological challenges, associated to integrating noisy/incomplete data, multiple conditions, prior knowledge, latent variables, or dealing with high dimensionality, unobserved variable derivatives, candidate library design, and uncertainty quantification.
Upon these criteria, sparse regression generally outperformed symbolic regression, particularly when using Bayesian frameworks. Next, deep learning and large language models, further emerge as innovative tools to integrate prior knowledge, although their reliability and consistency need to be improved. While no single method addresses all challenges, we argue that progress in learning digital twins will come from hybrid and modular frameworks combining chemical reaction network-based mechanistic grounding, Bayesian uncertainty quantification, and the generative and knowledge integration capacities of deep learning. To support their development, we further propose a benchmarking framework to evaluate methods across all challenges.
\end{abstract}

\section{Introduction}
Recent technological advances have made increasingly available multi-type datasets documenting biological variables at multiple scales (subcellular, single cell, bulk, patient), over a dynamical time window, and under different experimental conditions. To handle the complexity of such data, several statistical and mathematical approaches have been proposed, from mechanism-based models of regulatory networks that enable straightforward validation procedures, to black-box statistical models (e.g., neural networks), which generally offer limited possibilities of exploring predictive features. Thus, medical research leans towards mechanistic models that represent the underlying biology and predict the molecular mechanisms driving observed phenotypes.  Such \textit{digital twins} of the patients or biological systems constitute a powerful tool for the identification of disease-targetable vulnerabilities and personalized therapies. However, their design, i.e, the discovery of their governing equations, may be highly time-consuming for the human modeler as it requires integrating multi-type datasets with prior knowledge of various forms; this motivated the development of \textit{automatic model learning approaches}. 

Data-driven model inference was first developed for physics and engineering and allowed successful recovery of governing equations for systems that are low-dimensional, fully observable, and grounded in well-established mechanistic laws. Indeed, these domains benefit from high-quality time-resolved data with high signal-to-noise ratio and rich prior knowledge that guides model discovery. In sharp contrast, the knowledge in Biology is often partial or qualitative, data is noisy, sparsely sampled, and heterogeneous, and many relevant components remain unobserved. These challenges hinder the direct application of methods originally developed for physical systems to biological contexts.

Model learning for Biology began with intracellular network discovery, which aims to only identify gene interactions without capturing the system dynamics, a common example being Gene Regulatory Network (GRN) inference from transcriptomics ~\cite{marku_time-series_2023}. However, Biology may be better represented by quantitative models describing time-varying intracellular events using nonlinear ordinary differential equations (ODEs), where state variables are genes, proteins, or drug species undergoing chemical reactions governed by specified laws and rate constants~\citep{machado_modeling_2011}. Such a mathematical framework has been successfully applied to dynamically represent biological systems or drug pharmacokinetics-pharmacodynamics (PK-PD), thus populating the fields of systems biology and systems pharmacology~\citep{kitano_computational_2002}. Learning an ODE model implies inferring both i) the mathematical formulations of species interactions (e.g., law of mass action, Michaelis-Menten~\cite{keener_mathematical_2008}) and ii) the magnitude of the reactions (e.g., rate or enzymatic constants). Traditionally, this process is carried out manually: the modeler reviews the literature, formulates equations, and fits parameters to data. However, data collection and integration may be time-consuming, and such an approach is only feasible when the reaction network is known, which may be the central question of the project. Moreover, manual model inference presents reproducibility issues, as different modelers may yield different models. Hence, there is a need for a systematic approach to mechanistic model learning. 

In this review, we focus on methodologies inferring ODE-based digital twins from time-resolved biological datasets, which may document different cell types/patients and conditions (e.g., gene knockdown, drug treatment). We categorized methods as either symbolic or sparse regression algorithms~\citep{north_review_2023} and recapitulated them in a ShinyApp \footnote[1]{\url{https://u1331systemspharmacology.shinyapps.io/model_learning_review/}}. Emphasis was put on methods amenable to discovering complex systems and on innovative approaches combining classical regression with artificial intelligence (AI) tools~\citep{ghadami_data-driven_2022}. We identified four experimental and four methodological challenges currently associated with automatic learning of digital twins for Biology and Medicine, reported the performance of the most relevant methods and discussed the strategies to further manage them.

\section{Current approaches to data-driven discovery of digital twins}

\subsection{Problem Formulation}\label{sec:pb}

We are interested in inferring both the structure and the quantitative dynamics of an ODE-based model representing a biological system, capturing not only species interactions but also their precise mechanisms, and associated mathematical terms and parameters. 
Let us consider a biological system involving $m$ species, whose concentration vector $\y \in \R_+^m$ is experimentally observed on a discrete time grid provided as a dataset $\mathcal{D} = \{(t_j, \y_j)\}_{j = 1}^n$. The dynamics of the state variables $\x(t)\in \R_+^m$ are modeled through the following system of ODEs parametrized by $\theta\in \R_+^s$ :
\begin{equation}
    \left\{\begin{array}{l}
                            \begin{aligned}
                                \dot{\x}(t) &= \ff(\x(t),\theta ) \\
                            \x(0) &= \x_{0}\\
                            \y_j &= \x(t_{j}) + \bm{\varepsilon}_j
                       \end{aligned}\end{array}\right.
\label{eq:initpbem}
\end{equation}
where $\bm{\varepsilon}_j$ is the measurement error at observation time $t_j$. The key goal of model inference is to learn the vector field $\ff : \R^m_+, \R_+^s \to \R^m$ and parameters $\theta$ that best fit the provided datasets. 
Crucially, we seek to obtain a \emph{mechanistic} estimator $\hat{\ff}$ of $\ff$, composed of biologically grounded terms, typically derived from the mass action law, Michaelis-Menten, or Hill kinetics.

\subsection{State-of-the-Art of Digital Twin Inference from Temporal Data}\label{sec:state_of_the_art}

We reviewed existing computational methodologies inferring both structure and parameters of dynamical models from biological time-series data, and identified two major approaches. First, \textit{symbolic regression} (Section~\ref{sec:symr}) explores the space of mathematical expressions built from basic operators (e.g., $+, -, \times, \div, \exp, \log$) to identify models that best fit the data. Thus, it offers a flexible framework, not assuming any predefined model structure, but at the cost of a high computational burden. In contrast, \textit{sparse regression} (Section~\ref{sec:sr}) starts from a structured library of candidate functions (e.g., polynomial, trigonometric) and infers a minimal subset of terms using sparse regularization (e.g., Lasso). This yields parsimonious, interpretable models with greater efficiency when the library is appropriate. We describe hereafter each of these approaches that provide complementary tools for learning mechanistic models.

\subsubsection{Symbolic Regression}\label{sec:symr}

Unlike traditional regression, which aims to estimate from data the parameters of a user-defined model, symbolic regression encompasses a family of methods that simultaneously infer the model structure and parameter values that best describe a dataset $\mathcal{D}$~\citep{augusto_symbolic_2000, schmidt_distilling_2009}. Initially developed for static and low-dimensional problems, symbolic regression has since been extended to time series and dynamical systems, aiming to identify the functional form of the vector field $\ff$ in the ODE system (\ref{eq:initpbem}) directly from observations~\citep{dascoli_odeformer_2023, grigorian_learning_2025, galagali_exploiting_2019}. In the discovery of a $m$-sized system, symbolic regression seeks $m$ symbolic expressions $\{\hat{f}_l\}_{l=1}^m$ such that $\dot{x}_l \approx \hat{f}_l(\x)$ for each species $x_l$, minimizing a loss between model-predicted and data-derived derivatives. Most methods optimize such function independently for each species, thus enabling parallel computations, but making it difficult to enforce principles linked to variable interactions such as conservation laws.

\begin{figure}[h!]
    \centering
    \includegraphics[width=0.50\textwidth]{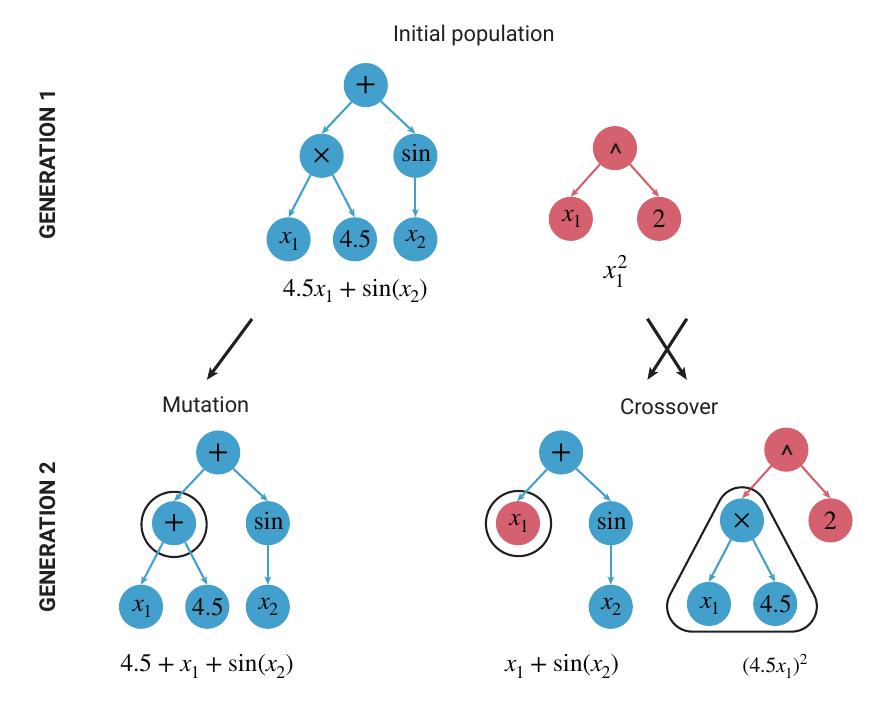}
    \caption{\textbf{Schematic of a typical genetic programming workflow for symbolic regression.} Candidate expression trees of the current iteration are used to generate the ones for the next generation through, for example, mutations or crossovers. Model selection is then guided by the goodness of fit to data, and only the top performers are passed to the subsequent iteration.}
    \label{fig:genetic_programming}
\end{figure}

These methods typically rely on \textit{expression trees} whose nodes are either a state variable, a constant or a basic operator, selected in a user-defined set  (e.g., $+, -, \times, \div, \exp, \log$) and whose vertices indicate the link between those items leading to explicit mathematical formulas (Figure~\ref{fig:genetic_programming}). The algorithms iteratively optimize the populations of candidate expression trees for each state variable independently. At iteration $i$, the $n_i$ candidate functions for estimating $x_l$ derivative, $\mathcal{P}_{i}^l = \left\{ \hat{f}^{(i)}_{1}, \hat{f}^{(i)}_{2}, \dots, \hat{f}^{(i)}_{n_i} \right\}$, undergoes perturbations, such as mutations and crossovers for Genetic Programming-based methods, to generate new expressions, which are then evaluated against the data, and only top performers are retained to form $\mathcal{P}_{i+1}^l $ (Algorithm~\ref{algo:symbolic}). Candidate expressions are typically refined \emph{via} evolutionary algorithms, Monte Carlo tree search, probabilistic grammars, or neural-symbolic models~\citep{la_cava_contemporary_2021}. 

Importantly, the general symbolic regression problem is considered NP-hard due to the exponential growth of the search space when increasing the number of variables and the considered operators, leading to high computational demands ~\citep{virgolin_symbolic_2022}. Moreover, traditional versions struggle to capture variable inter-dependencies as they learn equations separately for each variable, limiting their ability to infer coupled systems. Another limitation lies in the fact that symbolic regression may overfit the data without suitable regularization or selection strategies~\citep{de_franca_alleviating_2023}.

\begin{algorithm}[h!]
\caption{\textbf{General Symbolic Regression Algorithm}, encompassing most current approaches (evolutionary, neural, grammar-based, or constraint-guided), outlined for a 1-D problem. It presents the iterative process of symbolic regression through candidate populations initialization, selection, generation, and optional post-processing.}
\label{algo:symbolic}
\KwIn{Data $\mathcal{D}$, Symbolic expression set $\mathcal{F}$ or grammar $\mathcal{G}$, maximum iterations $N$, hyperparameter $\lambda$}
\KwOut{Best symbolic model estimator $\hat{f}$ with adjusted parameters $\theta^*$}

\vspace{1em}

\textbf{Step 1: Initialize the candidate models population $\mathcal{P}_{0}$:}\\
\Indp
\If{Using genetic programming or evolutionary methods}{
    Generate random initial population from $\mathcal{F}$
}
\ElseIf{Using neural or LLMs}{
    Sample candidate models from pretrained decoder or prompts using $\mathcal{F}$
}
\ElseIf{Using grammar-based methods}{
    Sample candidate models using predefined grammar $\mathcal{G}$
}
\ElseIf{Using other search strategies}{
    Build expressions by combining input variables and basis functions $\mathcal{F}$, guided by mathematical constraints.
}
\Indm

\For{$i = 0$ to $N$}{
  \textbf{Step 2: Optimize numerical parameters}\\
  \Indp
  \For{$k=1$ to $Card(\mathcal{P}_i)$}{
      Fit free parameters of $\hat{f}^{(i)}_k \in \mathcal{P}_i$ using numerical optimization (e.g., least-squares), yielding the best-fit parameter set $
      \theta^{*(i)}_k = \arg\min_{\theta^{(i)}_k} \; L\!\left(\hat{f}^{(i)}_k(\cdot;\theta^{(i)}_k), \mathcal{D}\right).$}
  \Indm
  
  \textbf{Step 3: Select promising models}\\
  \Indp
  \For{$k=1$ to $Card(\mathcal{P}_i)$}{
      Compute optional complexity score $C\left(\hat{f}^{(i)}_k(\cdot; \theta^{*(i)}_k)\right)$ (e.g., Minimum Description Length or Pareto filtering) \\
      Compute performance score using the minimized loss from Step 2: \[J^{(i)}_k \;=\; L\!\left(\hat{f}^{(i)}_k(\cdot; \theta^{*(i)}_k), \mathcal{D}\right) \;+\; \lambda\, C\!\left(\hat{f}^{(i)}_k(\cdot; \theta^{*(i)}_k)\right)\]
  }
  \hspace{2em}Retain the subset $\mathcal{S}_i \subset \mathcal{P}_i$ of top-performing candidates with the smallest $J^{(i)}_k$ 
  \Indm

  \textbf{Step 4: Generate new candidate models}\\
  \Indp
  \If{Using genetic programming}{
      Apply mutation and crossover on current candidates
  }
  \ElseIf{Using neural models or LLMs}{
      Generate models via decoding or prompting
  }
  \ElseIf{Using grammar-based methods}{
      Sample from grammar rules
  }
  \Indm
  
  \textbf{Step 5: Form new population $\mathcal{P}_{i+1}$}\\
  \hspace{2em}Combine selected and newly generated models\
}

\textbf{Step 6: Final optional post-processing}\\
\hspace{2em} Re-optimize parameters, simplify expressions\

\textbf{Return final model:}  $\hat{f}(.;\theta^*) \;=\; \arg\min_{k \in \{1,\dots,n_N\}} \left[ L \left( \hat{f}_k^{(N)}(\cdot; \theta^{*(N)}_k), \mathcal{D} \right) + \lambda\, C \left( \hat{f}_k^{(N)}(\cdot; \theta^{*(N)}_k) \right) \right]$ 
\end{algorithm}

\begin{figure}[h!]
    \centering
    \includegraphics[width=\textwidth]{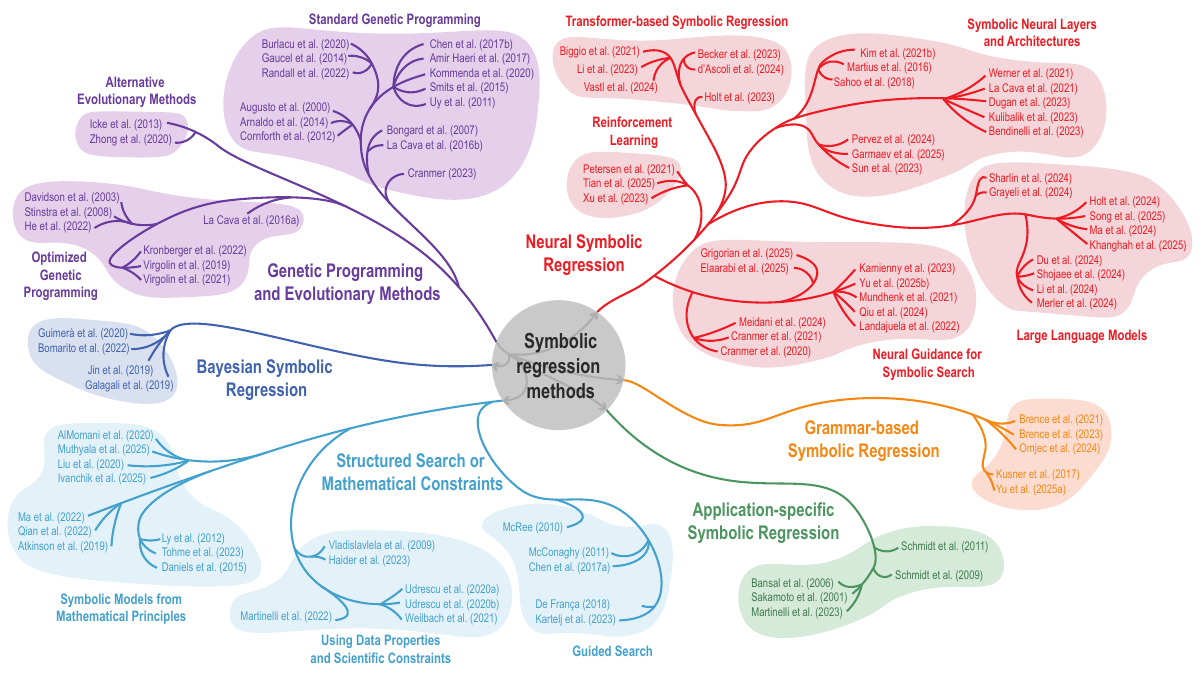}
    \caption{\textbf{Overview of symbolic regression methods.} This tree diagram categorizes existing approaches based on their core modeling principles, such as Genetic Programming and Evolutionary Methods (\cite{burlacu_operon_2020, gaucel_learning_2014, randall_bingo_2022, augusto_symbolic_2000, arnaldo_multiple_2014, cornforth_symbolic_2012, chen_feature_2017, amir_haeri_statistical_2017, kommenda_parameter_2020, smits_pareto-front_2005, uy_semantically-based_2011, bongard_automated_2007, la_cava_inference_2016, cranmer_interpretable_2023, icke_improving_2013, zhong_multifactorial_2020, davidson_symbolic_2003, stinstra_metamodeling_2008, he_taylor_2022, kronberger_shape-constrained_2022, virgolin_linear_2019, virgolin_improving_2021, la_cava_epsilon-lexicase_2016}), Neural Symbolic Regression (\cite{petersen_deep_2021, tian_interactive_2025, xu_rsrm_2023, biggio_neural_2021, li_transformer-based_2022, vastl_symformer_2024, becker_predicting_2023, dascoli_odeformer_2023, holt_deep_2023, kim_integration_2021, martius_extrapolation_2016, sahoo_learning_2018, werner_informed_2021, la_cava_contemporary_2021, dugan_occamnet_2023, kubalik_toward_2023, bendinelli_controllable_2023, pervez_mechanistic_2024, garmaev_nomto_2025, sun_pisl_2023, sharlin_context_2024, grayeli_symbolic_2024, holt_data-driven_2024, song_LLM-Feynman_2025, ma_llm_2024, khanghah_large_2025, du_large_2024, shojaee_llm-sr_2024, li_MLLM-SR_2024, merler_in-context_2024, grigorian_learning_2025, elaarabi_adaptive_2025, kamienny_deep_2023, yu_symbolic_2025, mundhenk_symbolic_2021, qiu_neural_2024, landajuela_unified_2022, meidani_snip_2024, cranmer_disentangled_2021, cranmer_discovering_2020}), Grammar-based Symbolic Regression (\cite{brence_probabilistic_2021, brence_dimensionally-consistent_2023, omejc_probabilistic_2024, kusner_grammar_2017, yu_grammar_2025}), Bayesian Symbolic Regression (\cite{guimera_bayesian_2020, jin_bayesian_2019, bomarito2022bayesian, galagali_exploiting_2019}), Structured Search or Mathematical Constraints (\cite{almomani_how_2020, muthyala_symantic_2025, liu_rode-net_2020, ivanchik_knowledge_2025, ma_evolving_2022, qian_d-code_2022, atkinson_data-driven_2019, ly_learning_2012, tohme_gsr_2023, daniels_automated_2015, vladislavleva_order_2009, haider_shape-constrained_2023, martinelli_reactmine_2022, udrescu_Ai-Feynman_2020, udrescu_Ai-Feynman-2.0_2020, weilbach_inferring_2021, mcree_symbolic_2010, mcconaghy_ffx_2011, chen_elite_2017, de_franca_greedy_2018, kartelj_rils-rols_2023}), or Application-specific Symbolic Regression (\cite{bansal_inference_2006, sakamoto_inferring_2001, martinelli_reactmine_2022, schmidt_automated_2011, schmidt_distilling_2009}).}
    \label{fig:map_symbolic}
\end{figure}

Figure~\ref{fig:map_symbolic} provides an overview of the symbolic regression methods we identified, which we grouped into six major families based on their core algorithmic strategies. First, \textit{genetic programming} (GP) methods iteratively evolve candidate mathematical expressions using operators such as mutation and crossover (Figure~\ref{fig:genetic_programming}), which are then selected over generations to minimize the prediction error on data, often incorporating structural constraints or complexity penalties. The earliest contribution of such type is by~\cite{bongard_automated_2007} and interestingly combines (i) partitioning, i.e., modeling each variable independently; (ii) automated model probing, suggesting experimental perturbations (e.g., changes in initial conditions or parameters) to diversify the inferred models; and (iii) snipping, i.e., simplifying expressions by pruning redundant or low-impact components to enhance interpretability. This method was successfully tested on the 3-D lac operon regulatory network in \emph{E. coli} with the best inferred model recovering 70\% of the target terms. Next, \textit{neural-symbolic regression} combines neural networks with symbolic modeling, via either analytical activation functions or symbolic decoders (i.e., neural modules that translate internal representations into symbolic expressions) that serve to generate candidate tree expressions at each iteration. More advanced strategies are also employed to constrain the symbolic search: reinforcement learning is used to efficiently explore the space of expressions, and neural networks can embed prior knowledge to guide the selection and refinement of candidate expressions. For example, ODEFormer~\citep{dascoli_odeformer_2023} treats equation recovery as a symbolic translation task: it encodes time series trajectories and directly generates the corresponding symbolic expressions using a transformer decoder.
It was evaluated on biologically relevant examples, including the SEIR epidemic model and the Lotka–Volterra system (4 and 2 variables, respectively), and showed strong performance and robustness even with noisy or limited data.
Once the framework is pretrained, inference of such low-dimensional models takes only a few seconds. \textit{Large Language Models} (LLMs) further extend this idea by generating candidate equation structures via prompts. Recent applications have shown strong results on biological systems, including tumor growth and pharmacokinetics~\citep{holt_data-driven_2024, du_large_2024, shojaee_llm-sr_2024}. Alternatively, \textit{grammar-based methods} rely on a formal specification of symbolic expression production rules as probabilities guiding the generation of new equation terms to be tested in the subsequent iteration. \cite{brence_dimensionally-consistent_2023} introduced grammar-guided sampling with unit-consistent constraints, enabling recovery of equations even with partial observability. Relevant to our current topic, \cite{omejc_probabilistic_2024} extended the method to ODE discovery from real-world data, and successfully recovered the dynamics of bacterial respiration and predator–prey systems, even with partial observability or low-frequency and noisy data. Next, \textit{Bayesian symbolic regression} infers posterior distributions of expressions by combining structural priors and data likelihoods. For instance, \cite{galagali_exploiting_2019} employed reversible-jump MCMC to infer both network topology and reaction rate parameters, and successfully identified nonlinear mechanisms of EGF–BRaf signaling. In a five-dimensional inference problem, their algorithm recovered the terms with a probability of about 0.75. Alternatively, \textit{ mathematical constraint-based methods} guide the model search by incorporating structural or domain-specific priors. Some methods leverage data properties or known scientific constraints (e.g., symmetries, conservation laws), others generate symbolic models from first principles (e.g., dimensional analysis), and a third group imposes structured search constraints based on equation syntax or consistency with trajectory behavior, reducing reliance on random exploration. Finally, of particular importance for this review, \textit{application-specific methods} tailor symbolic regression to biological or chemical systems. \cite{schmidt_automated_2011} refined ODE models of metabolic pathways by optimizing functional forms under biochemical constraints, such as stoichiometry and conservation laws. 
Reactmine~\citep{martinelli_reactmine_2022} sequentially identifies likely reactions, such as synthesis, degradation, or enzyme catalysis, based on preponderant changes in species concentrations.
The method inferred key known reactions in real-world datasets, one involving cell-cycle videomicroscopy,  and another, circadian regulation based on gene expression and biomarker data (3 and 5 variables, respectively). On a synthetic MAPK phosphorylation cascade (7 variables), 6 out of 7 underlying reactions were recovered.

Several benchmarks evaluate symbolic regression methods, starting from \cite{orzechowski_where_2018} who focused on low-dimensional algebraic expressions using noise-free data, and included only GP algorithms. Among those methods, EPLEX-1M~\citep{la_cava_epsilon-lexicase_2016} achieved the best generalization accuracy across 94 real-world datasets. Next,
SRBench~\citep{la_cava_contemporary_2021} expanded this first benchmark by incorporating neural methods and tests on dynamical systems, and revealed DSR~\citep{petersen_deep_2021} as the best-performing method in terms of accuracy and expression simplicity on both dynamical and algebraic tasks. SRBench++~\citep{de_franca_srbench_2024} extended the approach by including controlled noise and unified metrics, yet focusing on algebraic equations. In this last comparative study, PySR~\citep{cranmer_interpretable_2023} and Bingo~\citep{randall_bingo_2022} achieved the best overall harmonic trade-off between symbolic accuracy, robustness to noise, and expression simplicity, with uDSR~\cite{landajuela_unified_2022} showing perfect recovery on simpler tasks. Interestingly, PySR and DSR were also the best-performing methods at recovering chaotic systems, PySR being significantly faster and more robust on average ~\citep{gilpin_chaos_2021}. Next, ODEBench~\citep{dascoli_odeformer_2023} presents the most structured benchmark for ODE modeling, testing 63 systems, yet with only 12 of them involving more than 3 variables. ODEFormer~\citep{dascoli_odeformer_2023} consistently achieved the highest reconstruction and generalization accuracy across noise levels, while maintaining low inference time and expression complexity. 

Finally,~\cite{shojaee_LLM-SRBench_2025} introduced LLM-SRBench, a benchmark specifically designed to evaluate LLM-based symbolic regression methods. Instead of reusing equations, the authors generate synthetic one-dimensional equations by randomly assembling interpretable components from canonical biological mechanisms, like Hill functions, enriched with nonlinearities and structural diversity. This design ensures novelty and prevents trivial memorization by pretrained models. The benchmark includes 128 equations from various scientific domains, each paired with simulated time-series data, and evaluates symbolic coherence, numerical accuracy, and extrapolation. LLM-SR methods outperform traditional baselines for such metrics ~\citep{shojaee_llm-sr_2024}, but the best-performing inferred model achieved a symbolic accuracy of only 31.5\% . 

Overall, symbolic regression offers flexible formulations enabling the discovery of models faithfully representing the data, with the need for minimal prior knowledge, but currently tested only on models with a small number of variables. Existing benchmarks have enabled meaningful methods evaluation by highlighting important trade-offs between accuracy, interpretability, and generalization. Still, most benchmarks rely on synthetic, low-dimensional, and fully observed systems, often with known ground truths inside the search space, which can overstate performance.

\subsubsection{Sparse regression}\label{sec:sr}
Sparse regression is a powerful method for identifying governing equations of biological systems, leveraging the assumption that these systems typically have parsimonious representations, i.e., only a small subset of reactions significantly influences the dynamics~\citep{ouma_topological_2018}. 

Let us assume access to $Q$ longitudinal datasets $\Y^{(q)} \in \R^{n_q\times m}$, $q \in [1,Q]$, describing time-concentration profiles of the $m$ state variables evaluated at $n_q$ time points, potentially generated for different replicates, initial conditions or perturbations (e.g., gene knockout, drug exposure). These datasets are concatenated horizontally to yield $\Y \in \R^{n\times m}$, with $n = \sum_{q=1}^Q n_q$. 

The core idea of sparse regression is to approximate the vector field $\ff$ in Equation~\ref{eq:initpbem} as a linear combination of $p$ candidate functions from a library matrix $\boldsymbol{\Theta}(\Y) \in \mathbb{R}^{n \times p}$, where each column represents a potential term evaluated at each data time point. For instance, a typical library matrix may include constant, linear, polynomial, trigonometric, or nonlinear functions of the state variables:
\begin{equation}
    \bTheta(\Y) =
    \begin{bmatrix}
        | & | &  & | & | & & | & | & \\
        1 & \Y_{\bullet, 1} & \dots & \Y_{\bullet, m} & \Y_{\bullet,1}\Y_{\bullet,2} & \dots & \Y_{\bullet, m-1}\Y_{\bullet, m} & \cos \left(\Y_{\bullet,1}\right) & \dots\\
        | & | &  & | & | & & | & | &\end{bmatrix}.
\label{eq:thetaSINDy}
\end{equation}
The problem is then cast as solving
\begin{equation}
   \dot{\mathbf{X}} = \boldsymbol{\Theta}(\mathbf{Y}) \boldsymbol{\Xi} 
\label{eq:problem_SINDY}
\end{equation}
where $\dot{\mathbf{X}} \in \mathbb{R}^{n\times m}$ is the matrix of time derivatives of state variables and $\boldsymbol{\Xi} \in \mathbb{R}^{p \times m}$ contains the coefficients of each candidate term to be estimated. A sparse regression algorithm is applied to identify the smallest subset of active terms that best explains the system dynamics, thus enforcing that most entries of $\boldsymbol{\Xi}$ are zero, producing parsimonious but accurate and interpretable models. Overall, the common algorithm workflow for sparse regression methodologies involves: i) the estimation of derivatives or alternatively solving the ODE system, ii) assembling a candidate library of nonlinear terms, and iii) applying a sparsity-promoting regression scheme tailored to the desired prior knowledge or constraints (Algorithm~\ref{algo:sparse}).
This last step is typically formulated as minimizing a mismatch term plus a regularizer $\mathcal{R}(\bXi)$. The mismatch can be defined either at the derivative level, $\mathcal{L}(\bTheta(\Y), \hat{\dot{\mathbf{X}}}, \bXi)$, when $\dot{\x}(t)$ is estimated numerically, or at the trajectory level, $\mathcal{L}(\Y, \hat{\mathbf{X}}, \bXi)$, when reconstructed trajectories $\hat{\mathbf{X}}$ are obtained by integrating the candidate system.

A pioneering method in this field is the Sparse Identification of Nonlinear Dynamics (SINDy), introduced by \cite{brunton_discovering_2016}. It was the first to systematically apply sparse regression to the discovery of dynamical systems, using a predefined library of candidate nonlinear functions and enforcing sparsity through techniques such as Sequential Thresholded Least Squares, which iteratively alternates between least-squares fitting and hard thresholding to eliminate small coefficients. Although powerful and interpretable, the original SINDy method requires derivative estimation, making it sensitive to noise and mainly effective for low-dimensional, clean datasets. In terms of available implementation, the PySINDy Python package~\citep{kaptanoglu_pysindy_2021, silva_pysindy_2020} offers a user-friendly and actively maintained implementation of SINDy and related methods, including tools for constructing libraries or handling noisy data.

\begin{algorithm}[h!]
\caption{General Sparse Regression Workflow}
\label{algo:sparse}
\KwIn{Time series data $\Y$, hyperparameter $\lambda$}
\KwOut{Sparse model $\hat{f}$ describing system dynamics with adjusted parameters $\bXi^*$}

\vspace{1em}

\textbf{Step 1: Approximate derivatives or solve ODE system}\\
\Indp
\If{Approximated derivatives}{
    Estimate $\hat{\dot{\mathbf{X}}}$ using finite differences, splines, or neural approximators\
}
\ElseIf{Derivative-free formulation}{
    Approximate $\hat{\mathbf{X}}(\bXi)$ using integral, weak, or trajectory optimization formulation\
}
\Indm

\textbf{Step 2: Construct or refine candidate function library $\Theta(\mathbf{\Y})$}\\

\textbf{Step 3: Solve sparse regression problem}\\
\Indp
\If{Frequentist approach}{
Assume a data-fitting loss $\mathcal{L}$ and a sparsity-promoting regularizer $\mathcal{R}$  (e.g., $\ell_1$).

\Indp
\If{Approximated derivatives}{
   Solve: $\displaystyle \min_{\bXi} \mathcal{L}(\bTheta(\Y), \hat{\dot{\mathbf{X}}}, \bXi) + \mathcal{R}(\bXi)$ \
}
\ElseIf{Derivative-free formulation}{
    Solve: $\displaystyle \min_{\bXi} \mathcal{L}(\Y, \hat{\mathbf{X}}, \bXi) + \mathcal{R}(\bXi)$\
}
\Indm

}
\ElseIf{Bayesian approach}{
    Specify parameter prior $p(\bXi)$ (e.g., spike-and-slab, horseshoe)\\
    Compute posterior $p(\bXi|\Y) \propto p(\Y|\bXi)p(\bXi)$ , e.g., through MCMC \
}
\Indm

\textbf{Step 4: Model selection and validation}\\
\Indp
\If{Multiple models}{
    Select best model via cross-validation, AIC, Minimum Description Length or Pareto filtering\
}
\Indm

\textbf{Step 5: Optional refinement and interpretability enhancements}\\
\hspace{2em} Simplify expression, post-train parameters, analyze parameter sensitivity or identifiability\

\Return{$\hat{f}(.;\bXi^*)$}
\end{algorithm}

\begin{figure}[h!]
    \centering
    \includegraphics[width=\textwidth]{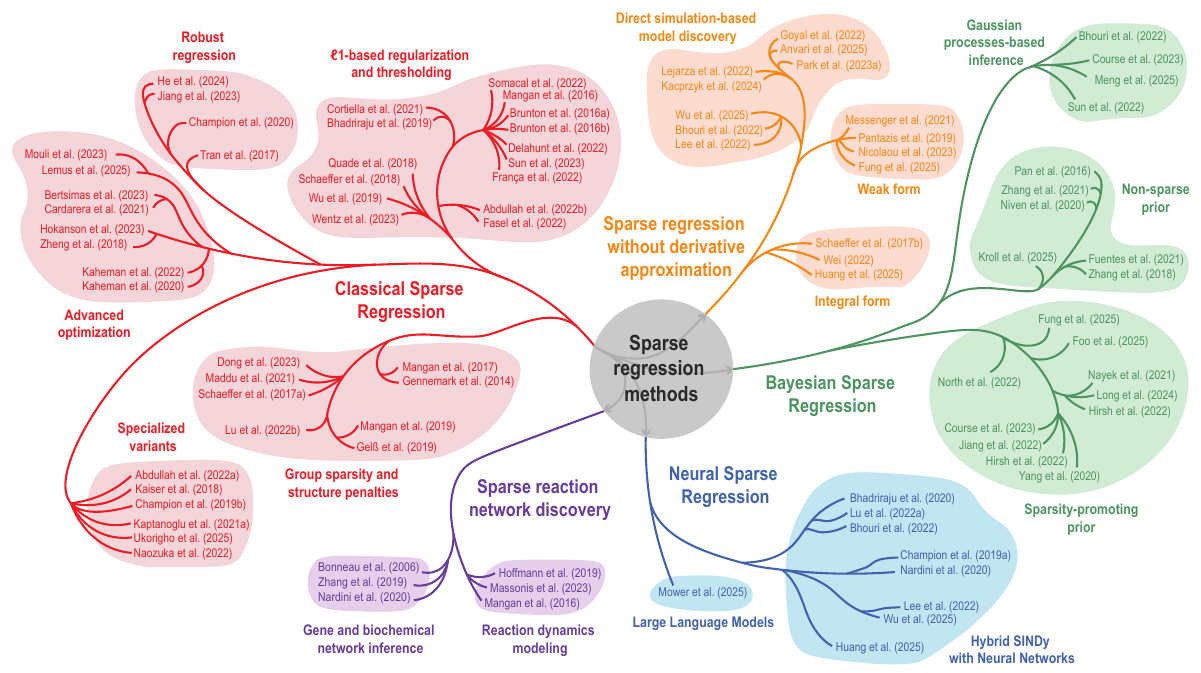}
    \caption{\textbf{Overview of sparse regression methods.} This tree diagram categorizes existing approaches based on their core modeling principles, including Classical Sparse Regression (\cite{somacal_uncovering_2022, mangan_inferring_2016, brunton_discovering_2016, brunton_sparse_2016, delahunt_toolkit_2022, sun_pisl_2023, franca_feature_2022, abdullah_modeling_2022, fasel_ensemble-sindy_2022, cortiella_sparse_2021, bhadriraju_machine_2019, quade_sparse_2018, schaeffer_extracting_2018, wu_numerical_2019, wentz_derivative-based_2023, he_sparse_2024, jiang_regularized_2023, champion_unified_2020, tran_exact_2017, mouli_metaphysica_2023, park_spreme_2023, lemus_multi-objectif_2025, bertsimas_learning_2023, carderera_cindy_2021, hokanson_simultaneous_2023, zheng_unified_2018, kaheman_automatic_2022, kaheman_sindy-pi_2020, abdullah_handling_2022, kaiser_sparse_2018, champion_discovery_2019, kaptanoglu_promoting_2021, ukorigho_competitive_2025, naozuka_sindy-sa_2022, dong_improved_2023, maddu_learning_2021, schaeffer_learning_2017, lu_sparse_2022, mangan_model_2017, gennemark_odeion_2014, mangan_model_2019,gels_multidimensional_2019}), Sparse Regression without Derivative Approximation (\cite{lejarza_data-driven_2022, kacprzyk_ode_2024, goyal_discovery_2022, anvari_implicit_2025, wu_data-driven_2025, bhouri_gaussian_2022, lee_structure-preserving_2022, messenger_weak_2021, pantazis_unified_2019, nicolaou_data-driven_2023, fung_rapid_2025, schaeffer_sparse_2017, wei_sparse_2022, huang_neuralcode_2025}), Bayesian Sparse Regression (\cite{bhouri_gaussian_2022, course_state_2023, meng_sparse_2025, sun_bayesian_2022, pan_sparse_2016, zhang_subtsbr_2021, niven_bayesian_2020, kroll_sparse_2025, fuentes_equation_2021, zhang_robust_2018, north_bayesian_2022, fung_rapid_2025, foo_quantifying_2025, nayek_spike-and-slab_2021, long_equation_2024, hirsh_sparsifying_2022, jiang_identification_2022, yang_bayesian_2020}), Neural Sparse Regression (\cite{mower_al-khwarizmi_2025, bhadriraju_operable_2020, lu_discovering_2022, bhouri_gaussian_2022, champion_data-driven_2019, nardini_learning_2020, lee_structure-preserving_2022, wu_data-driven_2025, huang_neuralcode_2025}), and Sparse Network or Reaction Discovery (\cite{bonneau_inferelator_2006, zhang_differential_2019, nardini_learning_2020, hoffmann_reactive_2019, massonis_distilling_2023, mangan_inferring_2016}).}
    \label{fig:map_sparse}
\end{figure}

Figure~\ref{fig:map_sparse} provides an overview of the reviewed sparse regression approaches, which we propose to group into five main families. First, \textit{classical sparse regression methods}, such as SINDy, infer governing equations from derivative approximations, using libraries of nonlinear candidate terms. This strategy has inspired many extensions aiming to improve SINDy robustness, interpretability, or model selection. For instance, SINDy-AIC~\citep{mangan_model_2017} outputs multiple models and selects the best ones using the Akaike Information Criterion to balance complexity and goodness-of-fit. It was shown to reliably recover true models for systems like SEIR epidemics (3 variables), for which itidentified the correct 6-term model. Next, \textit{sparse reaction network recovery} includes methods that are specifically amenable for biochemical or gene regulatory network structures from time series data. A representative method is Reactive SINDy~\citep{hoffmann_reactive_2019}, which constructs a library of candidate reactions following mass-action kinetics, enforcing chemical mass conservation that retrieved all true reactions of the \emph{E. coli} and of MAPK pathway networks (respectively, 3 and 9 variables). Alternatively, \textit{sparse regression methods without derivatives approximation} has been developed to bypass the sensitivity of numerical differentiation to data noise and sparse sampling. These methods reformulate the regression problem to avoid direct estimation of derivatives, using numerical integration or weak formulations. Next, \textit{Bayesian sparse regression} enables the use of sparsity-enforcing parameter priors such as spike-and-slab or horseshoe~\citep{piironen_sparsity_2017}, and help quantify uncertainty of inferred mathematical terms. For example,~\cite{jiang_identification_2022} apply a regularized horseshoe prior and MCMC sampling, and successfully recovered synthetic biological networks. Finally, \textit{neural sparse regression} techniques are increasingly used as they offer to leverage neural networks or LLMs for either estimating derivatives, modeling hidden components, or encoding prior knowledge to guide model search.

Despite many algorithmic variants of the popular SINDy framework, systematic benchmarking was lacking until~\cite{kaptanoglu_benchmarking_2023} introduced the first large-scale study using 70 three- or four-dimensional polynomial models of the standardized DYSTS database of chaotic dynamical systems. Their framework enables comparisons of five optimization strategies (i.e., STLSQ~\citep{brunton_discovering_2016}, Lasso~\citep{tibshirani_regression_1996}, SR3~\citep{zheng_unified_2018}, MIOSR~\citep{bertsimas_learning_2023}, weak SINDy~\citep{messenger_weak_2021}), under varying data noise levels. Notably, their results emphasize the strong performance of STLSQ and MIOSR and highlight the benefits of the weak SINDy formulation, both in terms of accuracy (low RMSE) and stability (low variance of RMSE across noisy experiments). However, this benchmark focuses solely on low-dimensional polynomial systems with full-state observability, which may not capture real-world challenges. Similarly, most studies introducing new model learning methods only compare their approach to one or two earlier baselines (typically STLSQ, Lasso) with limited hyperparameter tuning or statistical evaluation of the performances, reducing the reliability of comparisons. Critically, no benchmark yet evaluates sparse regression methods of all five categories or uses biologically realistic high-dimensional systems documented through irregular data.

Overall, sparse regression offers a flexible framework to automatically infer digital twins from biological time series. Compared to symbolic regression, it tends to produce more interpretable models due to the smaller number of terms ultimately selected by the methods. However, it may underfit when the true dynamics are not sparse or when the candidate function library lacks key terms, and can struggle with scalability in high-dimensional settings~\citep{bhouri_gaussian_2022, kaptanoglu_benchmarking_2023, mangan_model_2017}.


\section{Biological challenges}\label{sec:biological_challenges}
\begin{figure}[h!]
    \centering
    \includegraphics[width=1\linewidth]{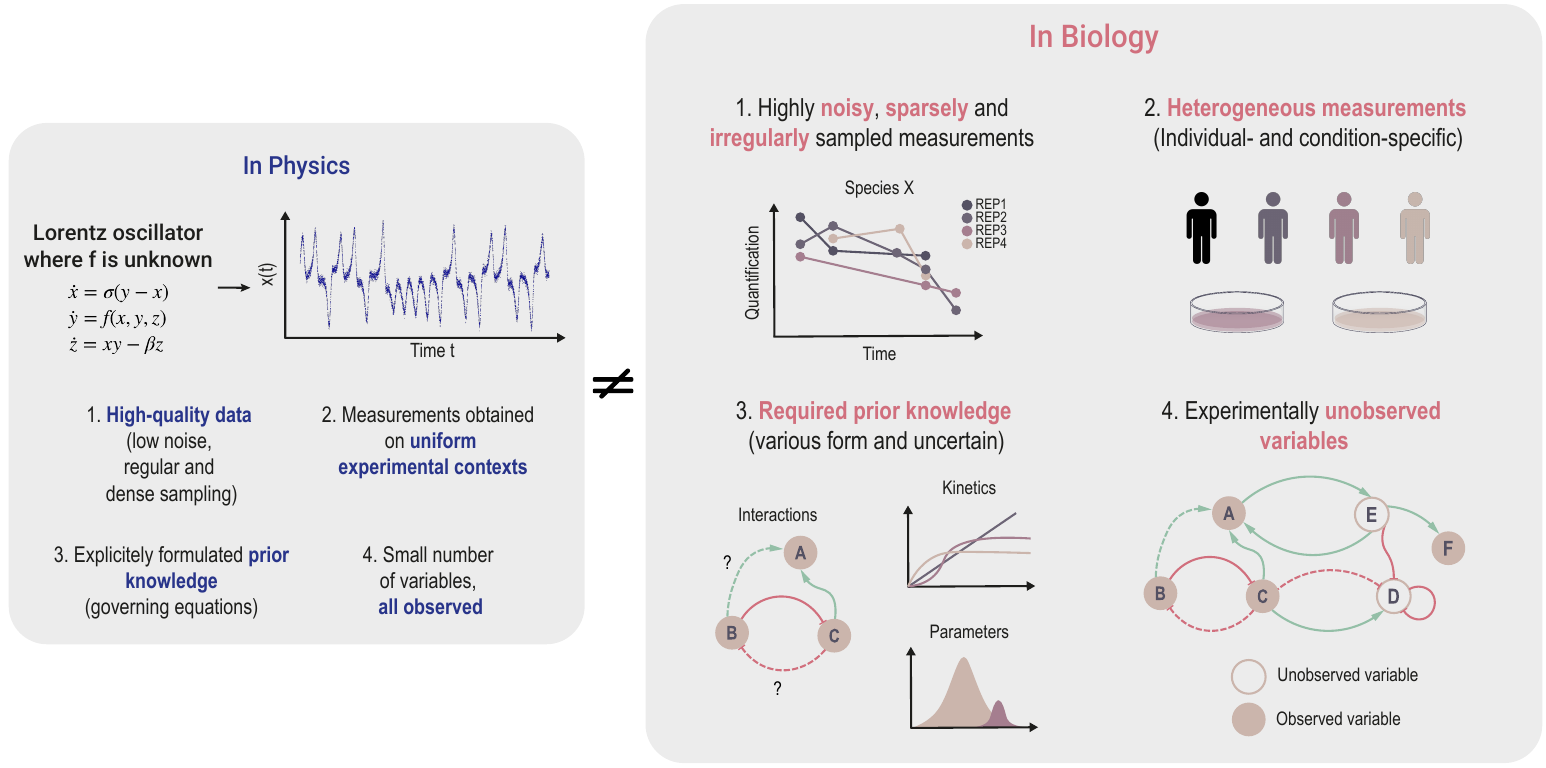}
    \caption{\textbf{Biology challenges in digital twin discovery.} As data-driven model learning was initiated for problems arising from physics (left panel), the resulting methods are not geared towards the unique challenges inherent to Biology (right panel).
    \textbf{Left:} Physical systems such as the Lorenz oscillator are typically documented by low-noise, densely and regularly sampled data available for a unique physical context, enabling full observability of a small number of variables. In addition, well-established prior knowledge already in the form of governing equations may further guide model search.
    \textbf{Right:} In contrast, biological data are often highly noisy, sparsely and irregularly sampled, complicating the reliable estimation of system dynamics.
    Measurements are frequently heterogeneous, varying across individuals and experimental conditions. Mechanistic prior knowledge (such as known species interactions, kinetic forms, or parameter ranges) is required to guide the model discovery and to constrain the model space in such low-data regime, but is often uncertain and of various types.
    Moreover, key variables can be experimentally unobserved, introducing hidden dynamics that further hinder model identification.
    These key differences highlight the unique challenges of applying model learning methods to biological systems.}
    \label{fig:pimlvsbiml}
\end{figure}

Unlike for physical systems or biological toy models commonly used for benchmarking model learning algorithms (e.g.Lotka-Volterra dynamics), the inference of real-world biological networks or digital twins involves specific challenges related to the high dimension of the systems resulting in a low-data regime, the complexity arising from inter-subject heterogeneity, and experimental and technical limitations (Figure~\ref{fig:pimlvsbiml}).

In the following, we discuss four main challenges in model learning; 1) the management of data irregularity and noise (Section~\ref{subsec:missing_data}), 2) the account of heterogeneous datasets across multiple conditions or individuals (Section~\ref{subsec:multiple_conditions}), 3) the integration of prior biological knowledge (Section~\ref{sec:pk}), and 4) the incorporation of unobserved variables in model discovery (Section~\ref{subsec:unobserved_variables}).

\subsection{Handling Noisy, Sparse, and Irregularly Sampled Biological Time Series}\label{subsec:missing_data}

In biological experiments, time series data are often irregular due to technical issues (detection limits, malfunctions) or intrinsic biological variability. This hinders the application of most model learning techniques, which usually assume that all variables are continuously and regularly observed. To address this challenge, two main strategies have been developed: (1) reconstructing the data before model search, and (2) directly integrating data uncertainty into the model inference process.

A first family of methods aims to reconstruct smooth, continuous trajectories from incomplete or noisy data. This preprocessing step enables the estimation of missing values with simple techniques such as spline-based interpolation and smoothing, low-pass or Savitzky-Golay filters, or kernel regression~\citep{wei_sparse_2022, bansal_inference_2006, sun_bayesian_2022, sun_pisl_2023, qian_d-code_2022, long_equation_2024, delahunt_toolkit_2022, lejarza_data-driven_2022}. More advanced methods rely on Gaussian processes, which combine smoothing, interpolation, and uncertainty quantification~\citep{atkinson_data-driven_2019, bhouri_gaussian_2022, course_state_2023, cranmer_interpretable_2023, meng_sparse_2025}, or on neural networks trained to denoise or interpolate sparse observations~\citep{nardini_learning_2020, qiu_neural_2024, yu_grammar_2025, wu_data-driven_2025}. Other strategies leverage an assumed or partially known model to reconstruct plausible trajectories. For instance, local Taylor approximations generate intermediate states by expanding around available data points (\citep{he_taylor_2022}) whereas trajectories may be generated from first-principle models (\cite{abdullah_handling_2022}). These denoising and reconstruction procedures may be crucial for improving the performance of model inference but must be manipulated with care to avoid erasing biologically-relevant information from data or creating artifacts.

A second category addresses data inconsistencies directly during model inference by adjusting the optimization procedure. For example, \cite{champion_unified_2020} proposed a \emph{trimming} functionality that automatically down-weights or removes data points with large residuals—often outliers or cases with missing replicates—allowing the algorithm to focus on reliable data. Such approach immediately raises the question of correctly tuning hyperparameters to avoid losing biologically-sound data points and associated information.
Following another rationale, \cite{omejc_probabilistic_2024} restrict the computation of the model error to observed components, at documented time points. \cite{north_bayesian_2022} further generalizes this by introducing a time-dependent observation mask that indicates which variables are measured at each time point, enabling flexible integration of partial trajectories. Other approaches directly address data uncertainty in the learning process such as the studies by \cite{hokanson_simultaneous_2023} and \cite{ kaheman_automatic_2022} that propose to jointly estimate a clean version of the data and the dynamic model.

\begin{tcolorbox}[
enhanced,
colback=myboxback_bio,
colframe=myboxframe_bio,
fonttitle=\bfseries,
title=Biological challenge 1 - Key takeaways,
coltitle=black,
boxrule=0pt,
arc=3mm,
drop shadow southeast,
left=2mm,
right=2mm,
top=1mm,
bottom=1mm,
toptitle=0mm,
bottomtitle=0mm
]
Overall, handling incomplete or noisy biological data may either be performed through an initial step of data reconstruction or denoising, or by an adaptation of the optimization procedure for model inference. Data smoothing and imputation are often critical for ensuring that inferred models reflect the underlying dynamics, especially in the presence of limited or corrupted observations. However, such procedures must be well adapted to the biological problem and raise the issue of hyperparameter tuning to avoid losing information or incorrectly transforming data. 
\end{tcolorbox}

\subsection{Leveraging Data in Multiple Experimental Conditions or Individuals}\label{subsec:multiple_conditions}

Biological datasets often include measurements in multiple cell lines, animal strains, or patients, and under varying experimental conditions such as gene knockouts or drug exposure. Fully exploiting this structure is essential for learning accurate and generalizable models in systems biology, where models are of high dimension and data is proportionally scarce. Yet, most model learning methods only process one condition at a time, limiting their ability to extract shared dynamics across experimental settings and to faithfully identify condition-specific variations. However, several recent methods have tried to address this limitation.

A promising strategy is to learn a shared model structure while allowing parameters to vary across individuals or conditions. \cite{schaeffer_learning_2017} propose a group-sparse regression method that enforces a common set of active terms across datasets while permitting coefficients to differ, capturing both invariant dynamics and inter-individual variability. Next, building on the same principle, INSITE~\citep{kacprzyk_ode_2024} first learns a global ODE model across all patients \emph{via} sparse regression, then personalizes it by re-estimating coefficients for each patient using their specific time series. \cite{park_spreme_2023} further unify this process with SpReME, which jointly learns a binary mask for the shared equation structure and condition-specific coefficients in a single optimization. These methods cleanly separate what is structurally conserved from what varies across conditions, enhancing robustness and interpretability.

Several frameworks leverage neural networks to discover or generalize shared dynamical structures. NeuralCODE~\citep{huang_neuralcode_2025} uses Automated Machine Learning (AutoML), i.e., automated architecture search, to design neural network architectures that separate shared from subject-specific parameters, effectively learning a common backbone of differential equations for biological processes consistent across patients, with personalized adjustments. In the same direction,~\cite{gui_discovering_2025} propose moving beyond explaining variability through parameter shifts alone \emph{via} invariant function learning, allowing both parameters and functional forms to vary across conditions. A neural network estimates the invariant component of the ODE, enabling, as a second step, symbolic or sparse regression to recover interpretable models for each subject. Alternatively, MetaPhysiCa~\citep{mouli_metaphysica_2023} casts dynamics learning as a meta-learning problem, identifying physical laws that generalize across experimental setups while efficiently adapting to new conditions, like unseen initial states or parameters. By combining causal inference and invariant risk minimization, it extracts robust, generalizable models even in extrapolation regimes. 
Complementing these approaches, OASIS~\citep{bhadriraju_operable_2020} pairs SINDy with a neural network that maps input conditions to ODE parameters. Local models are learned offline from condition-specific time series, while the network generalizes across them, enabling rapid adaptation to new regimes via online parameter prediction. Given their reliance on neural networks, these methods likely require larger amounts of data per patient to ensure accurate learning and generalization.

A different approach is proposed by~\cite{pantazis_unified_2019}, who use a multiple shooting approach: each trajectory (from different initial conditions or perturbations) is fitted separately, with the resulting equations being combined into a unified system where parameters are shared across trajectories. Then, sparse regression is used to recover the single dynamics that best fit all data. Finally, when datasets span heterogeneous regimes,  it may be necessary to learn multiple models. \cite{ukorigho_competitive_2025} address this with a competitive learning scheme that trains several models in parallel and softly assigns each data point to the best-fitting one based on prediction error, enabling discovery of distinct mechanistic regimes without prior labeling. However, such late integration strategies, where models are trained independently before being reconciled or selected, may be ill-suited to low-sample regimes where jointly leveraging all available data during training is crucial for robust generalization.

\begin{tcolorbox}[
enhanced,
colback=myboxback_bio,
colframe=myboxframe_bio,
fonttitle=\bfseries,
title=Biological challenge 2 - Key takeaways,
coltitle=black,
boxrule=0pt,
arc=3mm,
drop shadow southeast,
left=2mm,
right=2mm,
top=1mm,
bottom=1mm,
toptitle=0mm,
bottomtitle=0mm
]
Altogether, these approaches offer complementary strategies for handling multi-condition or multi-individual data, reflecting a growing need to explicitly model biological heterogeneity.
Methods that separate shared model structures with condition-specific parameters appear particularly promising, as they leverage all available data simultaneously during training. This makes them better suited for low-data biological settings, where pooling information is crucial to generalization. In contrast, neural network-based implementations often require substantial data per individual to be effective, which may not be realistic. Importantly, despite their potential, most methods remain to be tested on large-scale biological systems with multiple experimental settings.

\end{tcolorbox}

\subsection{Integrating Prior Biological Knowledge into Model Discovery}\label{sec:pk}

Prior knowledge may be defined as any supplementary information that helps constrain the otherwise vast and ill-posed problem of inferring equations governing a biological system from time series—such as known species interactions, kinetic mathematical forms, or plausible parameter values.

\subsubsection{Prior knowledge on chemical interactions}\label{sec:pki}

Prior knowledge about species interactions explicitly involves information regarding the possibility of a reactant $R$ becoming a product $P$. Such knowledge can be incorporated at the whole model level or at the reaction level, as described below.

Model-level priors encode global structural assumptions, such as \textit{sparsity} —favoring minimal interaction sets— or topological features like scale-free networks where most nodes have few connections and a few act as hubs~\citep{barabasi_emergence_1999, ouma_topological_2018}. These assumptions are commonly enforced in sparse regression frameworks (see Section~\ref{sec:sr}). For instance, SINDy~\citep{brunton_discovering_2016} uses $\ell_1$ regularization (LASSO,~\cite{tibshirani_regression_1996}), while SR3~\citep{champion_unified_2020} applies the sparsity penalty to an auxiliary variable rather than directly to the coefficients, improving optimization stability.
Bayesian approaches allow imposing flexible priors on parameters, such as the regularized horseshoe, which shrinks most coefficients to zero while allowing a few large ones, or the spike-and-slab, which explicitly models inclusion via a mixture of near-zero ('spike') and unconstrained ('slab') components~\citep{jiang_identification_2022, bhouri_gaussian_2022, nayek_spike-and-slab_2021}.
While accounting for such priors provides convex relaxations of the nonconvex model selection problem, Mixed-Integer Optimization (MIO) offers a more principled alternative by explicitly controlling term inclusion \emph{via} binary indicators that encode whether a given term is included (1) or excluded (0) from the model. As an important example of such method, MIOSR~\citep{bertsimas_learning_2023} formulates symbolic model discovery as a constrained optimization problem, enabling provably optimal recovery of ODE systems from noisy data.
Alternatively, Reactmine~\citep{martinelli_reactmine_2022} tackles model complexity through a tree-search algorithm that infers reactions sequentially, using a depth limit to enforce parsimony by assuming few reactions are sufficient to explain the data.

Next, chemical interaction-level prior information may be derived from curated biological databases, such as Protein-Protein Interaction networks (e.g., STRING~\cite{szklarczyk_string_2022}, SIGNOR~\cite{losudro_signor_2023}, HuRI~\cite{luck_reference_2020} and humanDB~\cite{peri_human_2004}), curated biochemical pathway databases like Reactome~\citep{milacic_reactome_2023}, KEGG~\citep{kanehisa_kegg_2024}, and BioCyc~\citep{karp_biocyc_2017}, and specialized resources like the Biological General Repository for Interaction Datasets (BioGRID~\cite{oughtred_biogrid_2021}) further documenting gene, protein, and chemical interactions. These resources provide structured knowledge that can be used to constrain or bias model selection toward biologically plausible structures. For example, Bayesian Reactive SINDy~\citep{jiang_identification_2022} allows specifying reaction-specific shrinkage priors to favor known interactions.
However, most existing inference frameworks lack standardized mechanisms for integrating biological priors, due to inconsistencies across databases, context-specific validity of interactions, and challenges in mapping qualitative interactions to quantitative ODE terms. As noted in~\cite{stock_leveraging_2025}, these obstacles—familiar from GRN inference— extend to ODE modeling, where the integration of information from biological databases remains an open and critical challenge.

\subsubsection{Prior knowledge on mathematical formulation or parameters of interaction kinetics}\label{sec:pkp}

Kinetic prior knowledge refers to assumptions or known forms of reaction kinetics, i.e., mathematical expressions of the reaction rates and associated parameter values. This can apply to the whole model or to specific interactions.

Systemic kinetic priors encode broad physical constraints such as conservation laws, dimensional consistency, and symmetries that reduce the candidate model space and enhance biological plausibility. Standard methods like SINDy ignore these, potentially producing unrealistic models (e.g., negative concentrations). Group sparsity methods can help by jointly selecting or discarding related terms~\citep{maddu_learning_2021}. Symbolic regression frameworks like ProGED~\citep{brence_dimensionally-consistent_2023, omejc_probabilistic_2024} enforce dimensional consistency, while NSRwH~\citep{bendinelli_controllable_2023} allows conditioning on user-defined hypotheses (e.g., symmetries), guiding expression generation. 
 
But ultimately, for Biology, it is essential to move beyond the general dynamical system formulation and adopt the chemical reaction network (CRN) formalism. CRNs offer a structured representation of biochemical processes by encoding known species interactions, canonical kinetic forms, and interpretable parameter constraints.

\begin{definition}
    A chemical reaction network (CRN) is a set of chemical reactions formally defined as a triple $(R, P, g_\theta )$, where $R$ (resp. $P$) is a multiset of $r$ reactants (resp. $p$ products) and $g_\theta: \R_+^r \to \R_+$ is a rate function over reactant concentrations specifying the reaction kinetics rates, parametrized by $\theta$. 
\label{def:crn}
\end{definition}

Crucially, casting ODE inference as CRN recovery imposes a strong structural prior: it narrows the model space to reaction-based dynamics, ensuring that the inferred vector field $\ff$ adheres to key mechanistic principles~\citep{fages_inferring_2015}. For instance, the stoichiometry of $(R, P)$ enforces mass conservation and structural constraints. 
As a result, $\ff$ is not only mathematically consistent—e.g., positive, smooth, and dimensionally correct—but also interpretable in terms of underlying biochemical processes. Such CRN-based formulations was applied in Reactmine~\citep{martinelli_reactmine_2022}, Reactive SINDy~\citep{hoffmann_reactive_2019}, its Bayesian variant~\citep{jiang_identification_2022}, and other studies~\citep{galagali_exploiting_2019, foo_quantifying_2025}.

Beyond global constraints, mathematical prior knowledge often exists at the species interaction level, such as known kinetics formulation for specific reactions. A key notion here is the \textit{reference model}—a predefined set of terms or equations subtracted from the estimated derivatives to perform model inference solely on residual dynamics~\citep{schmidt_distilling_2009, massonis_distilling_2023, martinelli_model_2021, martinelli_reactmine_2022}. 
Symbolic regression terms can also be restricted~\citep{randall_bingo_2022}. For example, \cite{schmidt_automated_2011} seeds symbolic regression with partial models, which are refined using data. Similarly, \cite{ivanchik_knowledge_2025} uses symbolic neural networks to bias term selection toward plausible forms.
At the parameter level, prior knowledge can be incorporated through informative Bayesian priors (e.g., Gaussian, horseshoe, spike-and-slab), derived from literature, experiments, or curated databases. These priors, used in methods such as Bayesian Reactive SINDy~\citep{jiang_identification_2022} and GP-NODE~\citep{bhouri_gaussian_2022}, help constrain parameter estimation, reduce uncertainty, and improve identifiability— which is again important in underdetermined or noisy biological settings.

\subsubsection{Foundation models-aided integration of prior knowledge}

Beyond direct inclusion of prior knowledge, the advent of Foundation Models (FMs) and LLMs offer new opportunities to incorporate more abstract forms of available information into model inference.
Pre-trained on large-scale, heterogeneous corpora like scientific texts, databases, and web contents,  these models internalize a wide range of domain knowledge, recognizing patterns and conventions that would otherwise require manual encoding.
In biology, for instance, specialized FMs like scGPT~\citep{cui_scgpt_2024} and scFoundation~\citep{hao_large-scale_2024} can generate single-cell transcriptomics data, as a result of their training on millions of gene expression profiles collected across diverse tissues and conditions, which allowed them to implicitly capture biological structures such as cell-type hierarchies and gene co-expression. Importantly for this review, such tools may assist tasks like model initialization or constraint suggestion.

Beyond that, synthetic datasets have also been exploited for pretraining model learning algorithms. ODEFormer is a Transformer—a neural network architecture designed to process sequences by dynamically weighting relationships between inputs—trained on millions of synthetic trajectories from randomly sampled ODEs~\citep{dascoli_odeformer_2023}.
It thus recapitulates common structural features of dynamical systems such as smoothness and variable dependencies. Such tool is especially useful for model inference from incomplete trajectory data as it can output symbolic equations for the underlying ODEs, which can serve as strong inductive priors. 
Similarly, Al-Khwarizmi~\citep{mower_al-khwarizmi_2025} integrates LLMs with symbolic regression, using prompts to ensure known dynamics are included in the learned model.
Interactive frameworks like Sym-Q~\citep{tian_interactive_2025}, LLM4ED~\citep{du_large_2024}, D3~\citep{holt_data-driven_2024}, and LLM-SR~\citep{shojaee_llm-sr_2024} further incorporate user feedback during the modeling process—for example, by allowing experts to accept, reject, or modify candidate equations—thereby steering the search toward models that are not only data-consistent but also aligned with established scientific understanding.

In addition, LLMs can assist in incorporating structured knowledge. The LLM-Lasso framework~\citep{zhang_LLM-Lasso_2025} exemplifies this by introducing adaptive, feature-specific penalties in sparse regression, guided by biological priors. Applied to SINDy, it assigns plausibility scores to candidate terms in the function library $\bTheta(\Y)$ using Retrieval-Augmented Generation~\citep{gao_retrival-augmented_2024}: relevant scientific documents (e.g., PubMed abstracts, pathway databases) are retrieved, and the LLM assesses each term’s biological relevance. For example, to evaluate the plausibility of the candidate term $x_i x_j$, the LLM receives a prompt consisting of literature mentioning $x_i$ and $x_j$ along with a query such as "Is the interaction between $x_i$ and $x_j$ biologically plausible?". This setup translates textual and database evidence into quantitative weights, effectively guiding interaction and kinetic priors by down-weighting implausible terms and promoting physically meaningful structures.
The framework also supports parameter priors by adjusting regularization using quantitative plausibility scores derived from retrieved literature that guide the strength of parameter shrinkage.

Yet, although promising, these methods come with serious limitations. First, simple baselines continue to outperform LLMs in certain predictive tasks such as perturbation prediction response~\citep{wentelerscpert, wong2025simple}.
Additional caveats also include limited biological interpretability, risks of propagating biases from training data, high computational costs, and the need to ensure outputs respect biological and physical constraints. Indeed, LLMs are particularly prone to hallucinations in underspecified settings~\citep{huang_survey_2025}, which are common in biology, making factual reliability a critical concern. Moreover, their effectiveness often depends on carefully crafted prompts, yet prompt engineering remains a manual, error-prone process~\citep{sharlin_context_2024}. Such limitations call for future benchmark studies, which are nowadays critically lacking in the field of digital twin design.

\begin{tcolorbox}[
enhanced,
colback=myboxback_bio,
colframe=myboxframe_bio,
fonttitle=\bfseries,
title=Biological challenge 3 - Key takeaways,
coltitle=black,
boxrule=0pt,
arc=3mm,
drop shadow southeast,
left=2mm,
right=2mm,
top=1mm,
bottom=1mm,
toptitle=0mm,
bottomtitle=0mm
]
Despite rapid progress, the integration of prior knowledge into biological model discovery remains underdeveloped and fragmented. While methods incorporating topological constraints and known mathematical terms or parameter priors into digital twin inference have recently been developed, this is often done in an \emph{ad hoc} or inflexible manner, and the automatic integration of large-scale structured biological databases remain challenging due to mapping difficulties and context sensitivity. Foundation Models and LLMs offer an intriguing new avenue for bridging these gaps, yet with limitations mainly due to their current lack of reliability that would deserve more in-depth evaluation.
\end{tcolorbox}

\subsection{Dealing with Latent or Unobserved Variables}\label{subsec:unobserved_variables}

In Biology, some elements may be unmeasurable due to technical or experimental constraints, resulting in undocumented variables that nonetheless influence the observed dynamics. Most current model learning methods assume full observability and cannot account for these hidden variables, limiting their real-world applicability. To address this, recent approaches include explicit latent trajectory inference, and neural reconstruction of hidden states.

A first family of methods incorporates user-defined latent variables and jointly optimizes their trajectories with observed ones. GP-NODE (\cite{bhouri_gaussian_2022}) and \cite{north_bayesian_2022} propose a probabilistic inference of both observed and hidden components within a Bayesian model. The problem formulation is similar to SINDy, using a candidate function library to express the system’s dynamics. The method then jointly infers the full-time evolution of all variables (including unobserved ones) and the sparse set of active terms in the equations, by fitting available observations. This task is performed in an end-to-end manner via either Hamiltonian Monte Carlo (HMC) or MCMC. Similarly, in a symbolic regression framework, \cite{omejc_probabilistic_2024} treats the initial conditions and parameters of user-specified unobserved variables kinetics as free parameters and optimizes them through numerical simulation to best match the observed outputs. For each candidate model generated from a probabilistic grammar, the full system of equations is simulated and compared to the available observed data. This allows the method to indirectly infer the influence of latent states on the measured dynamics, provided that the available data are sufficient to ensure practical identifiability.

An alternative is offered by the adaptive strategy of \cite{daniels_automated_2015}, which introduces hidden variables only when justified by predictive performance. Their method builds a hierarchical structure in an iterative manner: at each step, it evaluates whether adding a new latent variable reduces the prediction error on the observed data. If so, the variable is retained, otherwise, it is discarded. This approach allows automatically determining both the number and role of latent components, preserving parsimony while capturing hidden influences essential to accurate forecasting.

Another approach developed by \cite{somacal_uncovering_2022} proposes to regress higher-order derivatives of observed variables onto a dictionary of candidate functions, allowing the influence of hidden components to be encoded through derivative interactions. This leverages the fact that, under certain assumptions, unobserved states leave signatures in the derivatives of observed variables~\citep{takens_detecting_2006}.
In the same line,~\cite{martinelli_model_2021} models the indirect action of a variable $z$ through its time-integrated effect, $Z(t)= k\int_0^t z(s)\mathrm{d}s$, which assumes that its influence is mediated through an intermediate species, provided that the reaction follows the law of mass action. Interestingly, this enables modeling delayed - rather than instantaneous - regulatory effects when only upstream regulators are documented. Overall, indirect strategies like encoding latent influence \emph{via} higher-order derivatives or integral effects offer elegant workarounds but typically rely on strong assumptions holding in a limited number of cases.

A last class of approaches focuses on reconstructing latent states using neural networks.~\cite{lu_discovering_2022} propose a framework where a neural encoder maps temporal sequences of observed variables to estimate the unobserved components. From these reconstructed states, governing equations are then identified by fitting a sparse combination of predefined candidate functions. \cite{grigorian_learning_2025} similarly train a hybrid neural ODE model on the observed trajectories and apply symbolic regression to the latent part of the learned representation, yielding interpretable equations. These approaches combine the expressive power of neural networks with the interpretability of symbolic models, enabling recovery of hidden dynamics from partial observations. Yet, major challenges remain. Methods that explicitly infer latent trajectories often suffer from identifiability issues, making inference sensitive to initialization and priors. Neural reconstruction techniques, while flexible, may obscure mechanistic interpretability and can overfit in low-data regimes.

\begin{tcolorbox}[
enhanced,
colback=myboxback_bio,
colframe=myboxframe_bio,
fonttitle=\bfseries,
title=Biological challenge 4 - Key takeaways,
coltitle=black,
boxrule=0pt,
arc=3mm,
drop shadow southeast,
left=2mm,
right=2mm,
top=1mm,
bottom=1mm,
toptitle=0mm,
bottomtitle=0mm
]
An expanding toolkit is made available for incorporating unobserved variables in data-driven modeling, from Bayesian end-to-end approaches reconstructing hidden dynamics to neural networks learning latent temporal variables. However, such approaches raise the question of overfitting and model practical identifiability as the available data may not be sufficient to faithfully recover the equations and parameters related to unobserved variables. In this context, adaptive methods that introduce latent variables only when justified by improved prediction accuracy offer a promising way to control model complexity.
\end{tcolorbox}

\section{Methodological challenges}\label{sec:methodo_challenges}

\begin{figure}[h!]
    \centering
    \includegraphics[width=1\linewidth]{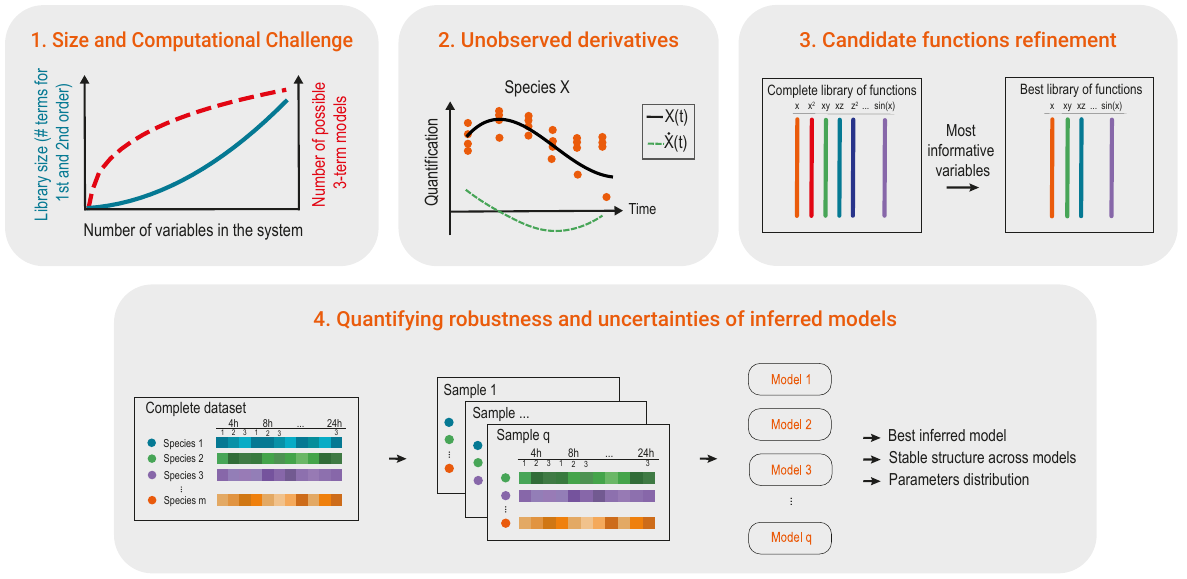}
    \caption{\textbf{The four main methodological challenges in digital twin discovery} (1) \textbf{High dimensionality of biological systems:} as the number of variables increases, the size of the candidate library and the number of possible models also grow, leading to major computational challenges, along with an increased need for data to ensure model identifiability. (2) \textbf{Handling unobserved derivatives:} when time derivatives are not directly available from data, one must rely on numerical approximation, alternative formulations of ODEs, or numerically solving the system. (3) \textbf{Selecting the candidate function library:} the choice of candidate functions shapes the search space and impacts the plausibility of inferred models. (4) \textbf{Quantifying robustness and uncertainties of inferred models:} identifying stable model structures across datasets and methodologies, and quantifying uncertainty in parameter estimates are crucial to ensure robust model inference.}
    \label{fig:challenges_methodo}
\end{figure}

Automatically learning digital twins from time series poses several methodological challenges, from common machine learning issues (e.g., optimization choices, regularization, hyperparameter tuning) to specific problems linked to inferring ODEs from biological data. Here, we highlight four key challenges: i) the high dimensionality of biological systems leading to method scalability and model identifiability issues, ii) the handling of unobserved time derivatives in a context of noisy and sparse measurements of state variables, iii) the choice of candidate function libraries for model learning, iv) the quantification of robustness and uncertainty of inferred models (Figure~\ref{fig:challenges_methodo}).

\subsection{High Dimension of Biological Systems leading to computational and low-data regime challenge}

One of the key differences between Physics and Biology lies in the models' dimension and complexity. Physical systems are typically low-dimensional and governed by compact, well-understood equations. In contrast, biological systems involve large networks of interacting species and greater structural complexity. Since most model learning methods have been initially developed for Physics, the unique challenges posed by biological systems' size have been largely overlooked.

These challenges are twofold. First, discovering larger systems requires high computational resources: as model dimension increases, sparse regression algorithms must handle rapidly growing candidate function libraries, while symbolic methods face a combinatorial explosion of possible expressions. For instance, performing sparse regression for a $10$-variable system and a function library $\bTheta(\Y)$ containing first- and second-order terms yields 65 possible terms; selecting up to 4 active terms per equation already produces over 722,000 possible combinations per equation. Second, high dimensionality requires more numerous and diverse datasets to constrain the model and avoid overfitting, which may be challenging considering the cost of biological experiments. However, with the rise of high-throughput technologies such as bulk and single-cell multi-omics profiling, CRISPR-based genetic or drug screening, collecting large-scale time-varying datasets is becoming increasingly feasible.

Nowadays, only a few model inference methods scale to moderately large systems. \cite{mangan_inferring_2016} developed a SINDy-based approach to infer a 7-dimensional yeast glycolysis model which required $\approx$900 temporal trajectories generated from distinct initial conditions, an experimental burden often difficult to meet in practice. Next, \cite{galagali_exploiting_2019} explored all 1024 subnetworks of a 10-dimensional system using a Bayesian framework, showing that their MCMC-based approach could shift between models involving one or two active pathways, demonstrating the utility of probabilistic methods in navigating medium-scale model spaces.

Other strategies aim to reduce the dimension of the problem.~\cite{gels_multidimensional_2019} use low-rank tensor decompositions to efficiently represent function libraries, exploiting multilinear structure to reduce complexity. However, while this scales to systems from 10 to 100 variables, it assumes that the dynamics admits a compact tensor-product structure, e.g., only involving functions like $\sin(x)$ and $\cos(x)$, which rarely holds in biological systems. 
More recently, \cite{sadria_discovering_2025} applied SINDy to gene programs extracted from single-cell RNA-seq data. By projecting expression profiles onto a 6-dimensional latent space, they enabled SINDy to operate on compressed representations of complex systems. However, the resulting dynamics remain confined to the latent space, lacking explicit correspondence to mechanistic biological variables, thus limiting biological interpretability.

\begin{tcolorbox}[
enhanced,
colback=myboxback_methodo,
colframe=myboxframe_methodo,
fonttitle=\bfseries,
title=Methodological challenge 1 - Key takeaways,
coltitle=black,
boxrule=0pt,
arc=3mm,
drop shadow southeast,
left=2mm,
right=2mm,
top=1mm,
bottom=1mm,
toptitle=0mm,
bottomtitle=0mm
]
While promising strategies exist, most methods still struggle beyond modest system sizes. A key challenge is to design scalable algorithms that retain interpretability and mechanistic accuracy in high dimension. This likely requires hybrid approaches that combine structural priors, modularity, and efficient inference to navigate vast model spaces without exhaustive search. Equally crucial is leveraging existing databases for systematically incorporating large-scale datasets like multi-omics profiling or genetic screens \label{sec:pk}, which provide rich perturbation data constraining dynamics recovery (Section~\ref{subsec:multiple_conditions}).
\end{tcolorbox}

\subsection{Approaches for Handling Unobserved Derivatives}\label{subsec:unobserved_derivatives}

A fundamental challenge in inferring systems of ODEs is the common unavailability of the vector field, as time derivatives are rarely experimentally measured. Consequently, methods fall into three broad categories: those that numerically estimate derivatives, those that approximate the candidate model solutions directly within the optimization procedure, and those that circumvent differentiation through reformulation of the problem.

In the first class, derivatives are estimated using either finite differences or smoothing methods like splines or Gaussian Processes~\citep{sun_bayesian_2022}, and then passed to the inference algorithm. Finite differences are fast but highly sensitive to measurement noise, which can dominate the derivative signal in biological data. Smoothing methods mitigate this by filtering out high-frequency fluctuations before differentiation, thereby producing more stable derivative estimates. However, this denoising step may also obscure sharp or transient biological features. Although straightforward, such two-step processes that fully decouple derivative estimation from model learning rule out the possibility of correcting for the bias of derivative approximation during the model inference phase.

Alternatively, recent approaches numerically solve the candidate models during the inference process, enabling their direct comparison with state variable measurements, in an end-to-end pipeline. For example,~\cite{he_sparse_2024} incorporate total variation regularization on the state trajectories during optimization. This encourages piecewise smooth solutions, which implicitly yield noise-robust derivatives without requiring their explicit computation. These regularized trajectories are then used internally for model fitting, tightly coupling trajectory estimation with parameter inference.
More expressive frameworks rely on Gaussian Processes or neural ODEs to jointly learn both the latent dynamics and observational noise. A prominent example is GP-NODE~\citep{bhouri_gaussian_2022}, which combines sparse identification of dynamics via SINDy, a differentiable neural ODE solver, and a Gaussian Process observation model. At each iteration, candidate dynamics (expressed as a sparse combination of nonlinear basis functions) are integrated via a neural ODE module to produce a continuous trajectory. This trajectory is then compared to the observed state variables time series and a Gaussian Process is used to model the mismatch: its mean is given by the neural ODE trajectory, and its covariance captures the residual uncertainty due to observation noise or model misspecification. 
These methods improve robustness but introduce challenges: higher computational cost, identifiability issues, and sensitivity to hyperparameters such as neural network depth or Gaussian process kernel choice.

 The last class of methods avoids derivative estimation by reformulating the inference problem. In integral formulations~\citep{goyal_discovery_2022, anvari_implicit_2025, schaeffer_sparse_2017, wu_data-driven_2025}, the ODE is rewritten as an integral equation, replacing differentiation with numerical integration. This makes the approach more robust to noise but introduces trade-offs such as increased computational cost and possible error accumulation over long trajectories.
Weak formulations~\citep{messenger_weak_2021, pantazis_unified_2019, qian_d-code_2022, nicolaou_data-driven_2023, fung_rapid_2025} extend this idea by projecting the dynamics onto a family of smooth test functions. This projection further enhances stability and noise resilience, and extensive benchmarking found the weak integral formulation to consistently outperform other strategies, even in noiseless cases—though their evaluation was limited to chaotic systems~\citep{kaptanoglu_benchmarking_2023, messenger_weak_2021}. However, design choices like selecting appropriate basis functions or integration intervals may significantly impact performance. 

\begin{tcolorbox}[
enhanced,
colback=myboxback_methodo,
colframe=myboxframe_methodo,
fonttitle=\bfseries,
title=Methodological challenge 2 - Key takeaways,
coltitle=black,
boxrule=0pt,
arc=3mm,
drop shadow southeast,
left=2mm,
right=2mm,
top=1mm,
bottom=1mm,
toptitle=0mm,
bottomtitle=0mm
]
Ultimately, the choice of strategy for handling undocumented derivatives affects both model accuracy and biological plausibility. Derivative approximation methods are simple and computationally efficient, but they are brittle under biological noise and risk irrevocably biasing subsequent inference. Noise may be better handled in end-to-end frameworks, approximating candidate model solutions to allow for direct data fit. Alternatively, reformulation-based approaches are also more robust to noise and better capture global trajectory structure. However, the latter two approaches typically come with trade-offs, including higher computational cost and sensitivity to hyperparameters.
\end{tcolorbox}

\subsection{Selection, Refinement or Generation of Candidate Function Libraries}\label{sec:candidatelibrary}

Candidate functions define the hypothesis space from which governing equations are inferred. In sparse regression, they form a predefined library of features; in symbolic regression, they emerge dynamically from combinations of variables, constants, and basic operators. While conceptually distinct, both approaches rely on suitable function sets to capture system dynamics. Overly limited candidate sets can hinder expressiveness, while overly large ones can overwhelm inference. To address this, various strategies offer to select, refine, or adapt candidate functions using data-driven or knowledge-based heuristics. These fall into two categories: properly selecting the candidate function library before model learning, or refining it during model discovery.

Optimized selection of candidate functions may be performed by Mutual information (MI), a widely used criterion that quantifies how much knowing one variable reduces uncertainty about another, thus helping to identify candidate functions strongly associated with the data we want to fit. Several model learning methods use MI for preliminary feature pruning  ~\citep{chen_feature_2017, muthyala_symantic_2025, song_LLM-Feynman_2025}. \cite{almomani_how_2020} further combine MI with entropy filtering, discarding near-constant variables before applying MI, ensuring selected features are both informative and sufficiently variable.
Other methods use structural constraints to reduce candidate sets. \cite{he_taylor_2022} apply Taylor expansions approximated from data to reveal properties such as separability or polynomial degree, and then use these properties as constraints in symbolic regression. \cite{zhang_robust_2018} apply dimensional analysis to retain only physically consistent terms in the library. \cite{bendinelli_controllable_2023} filter terms based on known symmetries (e.g., invariance under reflection or rotation), ensuring consistency with system properties such as conservation laws.

Next, candidate function engineering during model learning refers to the dynamical selection, transformation, or generation of features. 
Some sparse regression methods incorporate internal refinement mechanisms:~\cite{naozuka_sindy-sa_2022} use global sensitivity analysis to rank term importance during training and iteratively exclude the least performing candidate functions; and \cite{franca_feature_2022} combine Pearson correlation with an information-theoretic criterion to retain non-redundant, informative features in a two-step filtering process. \cite{bhadriraju_machine_2019} use stepwise regression with F-tests to iteratively retain statistically significant terms, while Dropout-SINDy~\citep{abdullah_modeling_2022} enhances robustness by randomly excluding subsets of candidate terms and aggregating submodels via median coefficients. 
Symbolic modeling methods may also involve candidate function refinement: GP-GOMEA~\citep{virgolin_improving_2021} applies linkage learning to evolve and recombine useful expression subtrees;~\cite{amir_haeri_statistical_2017} grow symbolic trees based on statistical heuristics favoring informative variables; and SyMANTIC~\citep{muthyala_symantic_2025} recursively expands expressions while enforcing low-complexity constraints to maintain interpretability.

Finally, deep learning techniques offer new ways to generate candidate functions by leveraging models pretrained on large corpora of equations and dynamical systems. Rather than relying on combinatorial construction from predefined operators, approaches such as~\cite{biggio_neural_2021, becker_predicting_2023, meidani_snip_2024, holt_deep_2023, dascoli_odeformer_2023} offer to map input data directly to symbolic equation terms. LLM-based methods extend this further by treating expression synthesis as a prompt-driven task, refining terms based on residuals of data fit~\citep{shojaee_llm-sr_2024, holt_data-driven_2024, du_large_2024}. In sparse regression, Al-Khwarizmi~\citep{mower_al-khwarizmi_2025} exemplifies this dynamic feature construction by incorporating visual inputs (e.g., annotated plots or schematics) into LLMs to build domain-specific candidate function libraries on the fly. These approaches shift candidate generation from static enumeration to a flexible, context-aware process embedded within the modeling pipeline. However, they face the typical limitations of LLM models, including a lack of reliability and biological relevance, and high dependency on prompt quality~\citep{sharlin_context_2024}. 

\begin{tcolorbox}[
enhanced,
colback=myboxback_methodo,
colframe=myboxframe_methodo,
fonttitle=\bfseries,
title=Methodological challenge 3 - Key takeaways,
coltitle=black,
boxrule=0pt,
arc=3mm,
drop shadow southeast,
left=2mm,
right=2mm,
top=1mm,
bottom=1mm,
toptitle=0mm,
bottomtitle=0mm
]
Taken together, these strategies demonstrate a shift from static candidate sets to adaptive, data-informed function spaces that evolve alongside the model inference. Selecting or refining candidate functions prior to or during model learning may not only reduce the hypothesis space and accelerate computation, but also facilitate the discovery of more accurate and interpretable models \citep{naozuka_sindy-sa_2022}. However, to our knowledge, no benchmark study has  quantified the actual gains in terms of model quality, robustness, and computational efficiency yet. Challenges remain as many feature selection methods rely on \emph{ad hoc} heuristics with limited theoretical grounding, and are based on a one-shot step rather than iterative workflows incorporating data fit residuals or expert feedback. 
While LLM-based approaches offer flexibility, they remain brittle, dependent on prompt quality, and are prone to producing overly complex or biologically implausible expressions~\citep{sharlin_context_2024}. 
\end{tcolorbox}

\subsection{Quantifying robustness and uncertainty of inferred models}\label{sec:uncertainty_quantif}
Digital twins are frequently applied in high-stakes domains like pharmacology and medicine, and ultimately serve for clinical decision-making, so that the accuracy of model predictions is of utmost importance. In such contexts, inferred models must not only reproduce observed data, but also remain robust to data perturbations (e.g., noise, subsampling, shifts in observation windows) and provide uncertainty quantification of parameter estimates, mathematical terms, or full equations. Crucially, the field’s historical focus on physical systems has led to the neglect of true biological robustness that must be retained in models. Biological networks typically involve multiple species connected through nonlinear, feedback-rich interactions (\cite[Figure 6]{kohn_molecular_1999};~\cite[Figure 4]{reinke_crosstalk_2019}) reflecting evolutionary pressure to maintain functions despite environmental noise, molecular fluctuations, and structural perturbations~\citep{whitacre_biological_2012, young_dynamics_2017}. Robustness arises through redundancy~\citep{hunter_understanding_2022} and dynamic compensation~\citep{kitano_biological_2004}, enabling systems to absorb perturbations while preserving essential behavior. Current inference frameworks largely ignore this resilience, yielding models that may fit data but fail to generalize across biological contexts. However, few methods integrate these aspects using either empirical approaches or Bayesian inference, as reviewed below.


Several empirical strategies assess the stability and confidence of inferred models by repeating the learning phase using synthetically perturbed datasets. \emph{Dataset bagging} (i.e., bootstrap aggregating) and sparse regression applied to each batch yields model ensembles, where robust terms are identified by their recurrence across runs~\citep{abdullah_handling_2022, fasel_ensemble-sindy_2022}.
Trajectory subsampling, i.e., repeating inference on different temporal segments, offers complementary insight into dynamic consistency under observational window shifts~\citep{wu_data-driven_2025, kartelj_rils-rols_2023, lejarza_data-driven_2022}. Next, for optimization-sensitive neural models whose results may significantly vary across runs, ensemble averaging can identify consistent dynamics across noisy replicates~\citep{dascoli_odeformer_2023} while multiple restarts with different initializations may provide a quantification of inter-run variability~\citep{nardini_learning_2020, grigorian_learning_2025}. Altogether, these empirical probes serve as practical tools for empirically assessing both model robustness and uncertainty.

As a second approach, Bayesian inference provides a principled framework to quantify uncertainty in both parameter values and model structure. At the parameter level, sparsity-promoting priors~\citep{nayek_spike-and-slab_2021, jiang_identification_2022, bhouri_gaussian_2022} yield posteriors that help identify which terms are consistently supported by the data (Section~\ref{sec:pkp}), thereby also yielding insights into robustness.
At the structural level,~\cite{jin_bayesian_2019} introduce a fully Bayesian symbolic regression framework where priors are placed on both expression trees and coefficients, enabling posterior inference over full equations via MCMC. This allows one to sample diverse candidate models and compute confidence scores of symbolic terms or motifs. Similarly,~\cite{dugan_occamnet_2023} and~\cite{galagali_exploiting_2019} propose strategies to approximate probability distributions of functional forms or reaction networks, respectively, thereby enabling structural uncertainty quantification. 

\begin{tcolorbox}[
enhanced,
colback=myboxback_methodo,
colframe=myboxframe_methodo,
fonttitle=\bfseries,
title=Methodological challenge 4 - Key takeaways,
coltitle=black,
boxrule=0pt,
arc=3mm,
drop shadow southeast,
left=2mm,
right=2mm,
top=1mm,
bottom=1mm,
toptitle=0mm,
bottomtitle=0mm
]
Despite recent advances, many methods struggle to establish the robustness of inferred models and standard tests are critically needed to assess model stability under realistic biological variations. 
Uncertainty quantification also faces key limitations, as many methods rely on heuristics like ensembling or dropout that lack formal grounding. Bayesian approaches offer more principled estimates but are computationally costly and scale poorly to large networks. Even when posterior distributions are obtained, they often suffer from overconfidence and poor identifiability. 

\end{tcolorbox}

\section{Synthesis on Best-performing Model Learning Methods for Biology}

Our review highlights the diversity and complementarity of existing sparse or symbolic regression methods for dynamic model discovery, and best performing methods are presented in two comparative tables (Tables~\ref{fig:table_symbolic} and~\ref{fig:table_sparse}).

\subsection{Strengths and limitations of existing symbolic regression methods in terms of identified challenges}\label{sec:synthesis}

\begin{table}[h!]
    \centering
    \includegraphics[width=0.85\linewidth]{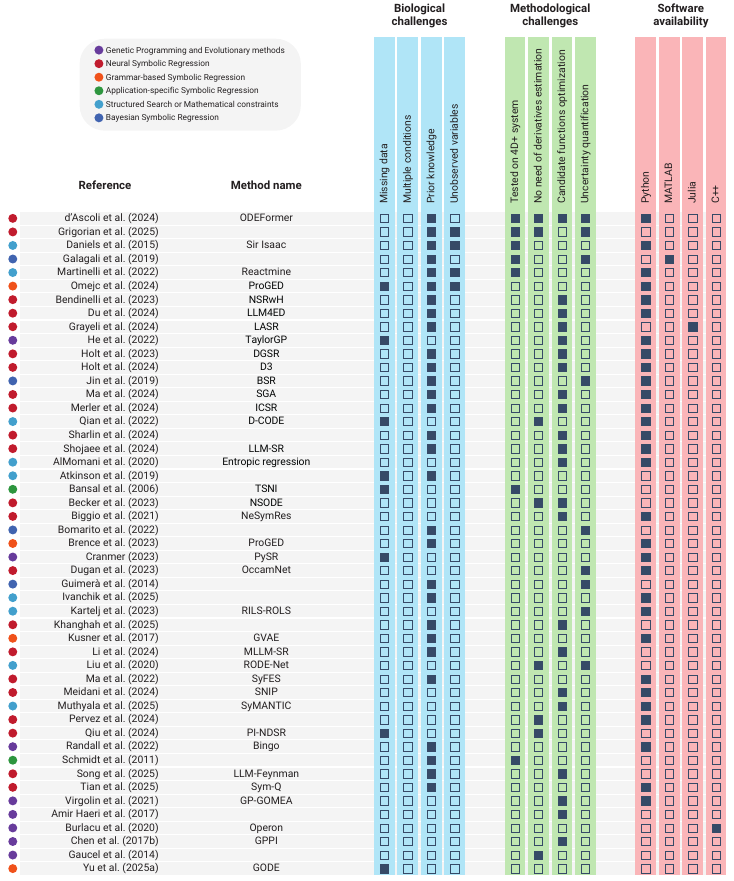}
    \caption{\textbf{Comparative overview of symbolic regression methods} selected for addressing key challenges in biological model discovery. Each row corresponds to a method, and each column indicates whether the method tackles specific biological challenges (missing data, multiple experimental conditions, prior knowledge integration, unobserved variables), or specific methodological challenges (avoids numerical derivative estimation, performs uncertainty quantification, has been tested on high-dimensional systems, or includes an optimization of the candidate function set). The table also lists the programming languages available for each method. Methods are ordered such that those addressing the largest number of challenges and providing an available implementation appear at the top. Colored disks indicate the symbolic regression category of the method. Methods with limited applicability to biological model inference were excluded for clarity.}
    \label{fig:table_symbolic}
\end{table}

Symbolic regression methods offer an interpretable framework for discovering compact analytical expressions, yet they struggle to meet the specific requirements of biological modeling as summarized in Table \ref{fig:table_symbolic}. Among the 49 best-performing symbolic regression methods reviewed here, only three of them address more than three challenges out of the eight identified ones. Eleven out of 49 algorithms handle either "Missing data" or "Unobserved variables" while only one method may incorporate both. In addition, none of the reviewed methods provides solutions for the challenge of integrating "Multiple experimental conditions". Similarly, symbolic regression methods largely do not meet key methodological criteria such as avoiding derivative estimation (8/49 methods) and uncertainty quantification (9/49 methods). Importantly, high-dimensional systems remain particularly challenging: only seven methods have been tested on systems with four or more state variables, and some studies report severe computational burdens in such settings \citep{cranmer_interpretable_2023, yu_grammar_2025}. These unresolved challenges are compounded by a structural limitation: most symbolic regression frameworks neglect explicit modeling of inter-variable dependencies, making most of them poorly suited for inferring coupled ODE systems. Nonetheless, symbolic regression approaches show remarkable efforts in the generation and selection of candidate functions (22/49 methods), as detailed in Section \ref{sec:candidatelibrary}. In parallel, 29 methods integrate prior knowledge, most often through pretrained neural networks or LLMs, offering complementary pathways for enhancing model discovery. 

Among existing methods, two neural-based frameworks by\cite{dascoli_odeformer_2023}  and \cite{grigorian_learning_2025}, respectively, stand out for their innovative contributions, addressing five out of the eight challenges identified in this review. \cite{dascoli_odeformer_2023} proposed ODEFormer, a Transformer-based model trained on synthetic ODEs, capable of recovering closed-form symbolic dynamics from a single noisy trajectory. On the other hand, \cite{grigorian_learning_2025} developed a hybrid method combining a neural ODE for latent trajectory inference with symbolic regression. Both methods successfully deal with the integration of prior knowledge, the applicability in high dimensions, unobserved derivatives, and uncertainty quantification, while solely ODEFormer includes an advanced mechanism for selecting candidate functions, and solely \cite{grigorian_learning_2025} can successfully recover partially observed systems. For ODEFormer, the inference of undocumented variables could be implemented via neural surrogates, following the approach of \cite{grigorian_learning_2025}. Of note, neither methods currently handle missing data, which may be addressed through mechanisms developed in other model learning studies: interpolation using splines, Gaussian processes, or neural reconstruction, as discussed in Section \ref{subsec:missing_data}. 
Next, neither methods allows for the integration of datasets from multiple experimental conditions. While direct incorporation into these symbolic architectures may be complex, a two-step strategy inspired by \cite{gui_discovering_2025} could offer a feasible and modular alternative, by first learning a neural invariant representation across conditions, and subsequently performing symbolic regression on the extracted features. Overall, these adaptations would help apply symbolic regression methods in more realistic and applicable biological modeling scenarios.

\subsection{Strengths and limitations of existing sparse regression methods in terms of identified challenges}\label{sec:synthesis}

As compared to symbolic regression approaches, sparse regression methods are generally more effective at addressing the biological and methodological challenges investigated in this review (Table \ref{fig:table_symbolic} and \ref{fig:table_sparse}). Indeed, among the 55 reviewed methods, 18 address more than three challenges (as compared to 3/49 for symbolic regression). Regarding biological issues, nearly half of them handles missing data (25/55 methods) or the integration of prior knowledge (22/55 methods), most often through prior distributions on parameters. In terms of methodological challenges, 23 methods include uncertainty quantification, typically through posterior distributions or bootstrap procedures (Section \ref{sec:uncertainty_quantif}), while 26 avoid the need for derivative estimation, which is a major advantage for mitigating noise. Overall, sparse regression more frequently addresses the challenges highlighted in this review and appears more adapted to learn biological systems as compared to symbolic regression. However, only a limited subset of frameworks may integrate multiple experimental conditions (8/55 methods) or unobserved variables (4/55 methods). Nevertheless, the solutions proposed for these challenges—discussed in Sections \ref{subsec:multiple_conditions} and \ref{subsec:unobserved_variables}— could inspire the development of more comprehensive approaches. Finally, only six methods incorporate candidate function selection or refinement, despite the fact that such procedures exist (see section \ref{sec:candidatelibrary}) and could be implemented in sparse regression approaches to greatly improve method performance \citep{naozuka_sindy-sa_2022}.

Among sparse regression approaches, the most effective methods that address the highest number of identified biological and methodological challenges were all Bayesian frameworks, demonstrating the superiority of such approach for model learning (\cite{bhouri_gaussian_2022, jiang_identification_2022, long_equation_2024, course_state_2023, foo_quantifying_2025, fung_rapid_2025, north_bayesian_2022, sun_bayesian_2022}).  The methods by \cite{bhouri_gaussian_2022}, \cite{jiang_identification_2022}, and \cite{long_equation_2024} emerge as the only pipelines addressing more than five out of the eight challenges considered in this review. All three methods i) embed interpolation or trajectory reconstruction techniques to mitigate the effects of data noise or missing values, ii) integrate prior knowledge through sparsity-enforcing parameter priors, iii) were validated on biological systems with at least 4 dimensions, iv) directly perform a numerical resolution of the ODE system to avoid the instability of finite differences-based derivative approximations, v) offer a quantification of model uncertainty through parameter posterior distribution. \cite{bhouri_gaussian_2022} go further by handling unobserved variables (see section \ref{subsec:unobserved_derivatives}). To further improve these methods, several directions emerge from the review. For handling multiple experimental conditions, Bayesian sparse regression is compatible with multi-condition model discovery as in \cite{schaeffer_learning_2017} and \cite{park_spreme_2023}. Another solution, particularly when larger datasets are available, is to adopt a two-stage pipeline using the neural representation learning approach of \cite{gui_discovering_2025} to capture invariant functions and apply sparse Bayesian inference on those learned functions to extract interpretable mathematical structures. Finally, regarding candidate function selection, most documented sparse regression methods would benefit from incorporating statistical criteria upstream or during inference, as discussed in Section \ref{sec:candidatelibrary}.

\begin{table}[h!]
    \centering
    \includegraphics[width=0.83\linewidth]{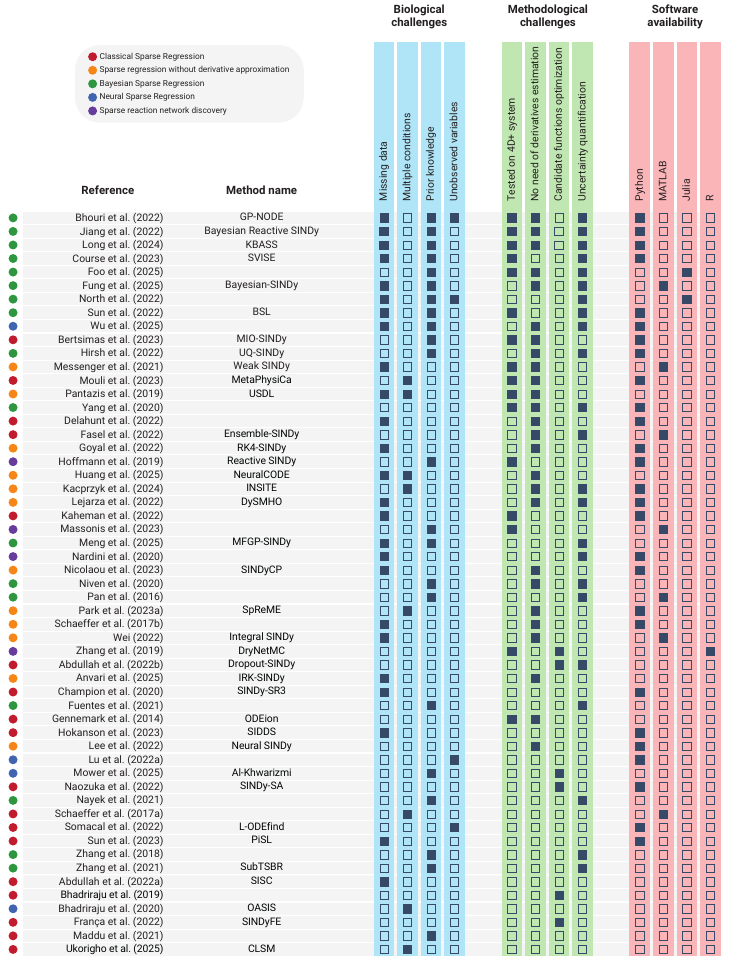}
    \caption{\textbf{Comparative overview of sparse regression methods} selected for addressing key challenges in biological model discovery. Each row corresponds to a method, and each column indicates whether the method tackles specific biological challenges (missing data, multiple experimental conditions, prior knowledge integration, unobserved variables), or specific methodological challenges (avoids numerical derivative estimation, performs uncertainty quantification, has been tested on high-dimensional systems, or includes an optimization of the candidate function set). The table also lists the programming languages available for each method. Methods are ordered such that those addressing the largest number of challenges and providing an available implementation appear at the top. Colored disks indicate the symbolic regression category of the method. Methods with limited applicability to biological model inference were excluded for clarity.}
    \label{fig:table_sparse}
\end{table}

\section{Discussion}

In this review, we surveyed recent advances in the data-driven inference of biological digital twins represented as mechanistic ODEs, focusing on two main approaches —symbolic or sparse regression— summarized in an interactive Shiny application\footnote[1]{\url{https://u1331systemspharmacology.shinyapps.io/model_learning_review/}}.
We assessed how these methods address four key biological challenges: noisy and irregular data, heterogeneity across conditions, incorporation of prior biological knowledge, and partial observability of system states. We also considered four methodological issues associated with scalability to high-dimensional systems, unobserved state variable derivatives, candidate function library design,  and model uncertainty quantification.
Overall, sparse regression emerges as more efficient than symbolic regression, particularly in Bayesian formulations, which excel at incorporating prior knowledge and quantifying uncertainty in both parameters and inferred models — crucial features in noisy, underdetermined biological settings. However, no inference method addresses more than six out of the eight challenges, which advocates for future developments.
Scalability to high-dimensional networks remains a main bottleneck, as candidate libraries grow combinatorially and data demands rise with model size.

Of utmost interest,
deep learning methods are increasingly combined with sparse and symbolic regression, resulting in improved performance.
In sparse regression, neural ODEs help denoise trajectories and capture latent dynamics before term selection~\citep{bhouri_gaussian_2022, wu_data-driven_2025}. In symbolic regression, hybrid neural ODEs blend neural approximators with domain constraints to guide expression search and improve interpretability~\citep{grigorian_learning_2025, bendinelli_controllable_2023}. These approaches complement foundation and large language models, which offer new ways to exploit prior knowledge from literature, structured databases, or expert input. 
As an important example, LLMs trained on ODE simulations can predict symbolic equations from incomplete trajectories, which can serve as an educated prior or for assigning plausibility scores to candidate library terms based on biological context.
However, their lack of explicit temporal modeling limits their ability to simulate dynamics, and challenges remain regarding their consistency, factual reliability, and bias, underscoring the need for rigorous evaluation and control.

In conclusion, while Bayesian frameworks and deep learning have pushed the field forward, critical challenges remain before digital twin discovery can be routinely deployed in biological applications. Addressing scalability, establishing rigorous benchmarks, and responsibly integrating AI tools such as LLMs and foundation models will be key to advancing our understanding of complex biological systems and translating dynamical modeling into clinically meaningful discoveries.

\section{Perspectives}

\subsection{Toward integrated and reliable frameworks}

Overall, our analysis shows that no single model learning method currently addresses all the challenges of biological systems, with real-world applicability being still limited.
Since each challenge has been addressed in at least one study, the key insight is that progress is likely to come less from entirely new algorithms than from integrating existing ones into flexible, modular frameworks.
To that end, we advocate for a CRN-based approach, which naturally encodes biological structure through stoichiometry, kinetics, and reaction directionality (Section~\ref{sec:pk}).
Effective frameworks should also prioritize reliability under adverse conditions—avoiding explicit differentiation, refining candidate functions, and rigorously quantifying model uncertainty.
Notably, few methods currently distinguish epistemic uncertainty arising from limited knowledge or model misspecification from aleatoric noise, limiting interpretability. Future work should assemble these components into general-purpose tools for dynamic model discovery in biology, a necessary step toward realizing the full potential of data-driven modeling in complex living systems.

Looking ahead, the most exciting opportunities lie in combining these approaches with emerging AI models. In this area, promising directions include (i) hybrid pipelines that integrate mechanistic priors with LLM-guided candidate generation, (ii) retrieval-augmented strategies that anchor outputs in curated sources such as pathway databases, and (iii) tools that translate expert input or experimental context into executable constraints, making models more adaptive and user-driven.

\subsection{The need for comprehensive benchmarks}

Our review also underscores the critical need for comprehensive benchmarks to objectively evaluate inference methods.
Algorithm evaluation mostly relies on synthetic data generated from a predefined model—corresponding to the ground truth to be recovered—which then serves as inputs for inference.
To ensure practical relevance, curated biological ODE systems from repositories such as BioModels~\citep{malik-sheriff_biomodels_2020} should be prioritized.
Moreover, alongside classic low- or medium-dimension models (e.g., Lotka-Voltera, yeast glycolysis, or MAPK signaling~\citep{mangan_inferring_2016, martinelli_reactmine_2022}), larger systems such as the circadian clock~\citep{hesse_mathematical_2021} must also be included, as they feature feedback loops and oscillatory behavior.
These curated models provide biologically plausible testbeds with known structure and parameters for assessing recovery performance, identifiability, and interpretability.

Next, efficient benchmarks must also probe the four biological challenges highlighted in this review.
(i) \textbf{To evaluate sensitivity to data quality limitations} such as noise, sparse sampling, and irregular time grids, benchmarks should systematically vary synthetic data parameters like noise level, observation frequency, and number of initial conditions. (ii) \textbf{To assess how methods handle biological variability}, benchmarks can emulate inter-subject or inter-condition differences by altering the model parameters while preserving the underlying network structure. For instance, reaction rates may be perturbed to reflect cell-type differences, or set to zero to simulate knockouts; initial concentrations can be varied to mimic genetic heterogeneity. (iii) \textbf{To probe the influence of prior knowledge}, one should vary both the quantity and reliability of available priors by gradually introducing partial reaction networks or known kinetic forms, or deliberately including misleading information such as spurious interactions. This reveals how methods balance data-driven discovery with prior constraints. (iv) \textbf{To test the ability to recover hidden dynamics}, benchmarks should vary the fraction of observed species, thus testing inference under partial observability. Since these challenges often interact, combining them will yield more realistic evaluations and enable challenge-aware comparison of inference methods.

\section{Methods}

\paragraph{Review methodology}
This review was conducted using a non-systematic strategy starting by querying PubMed and arXiv using combinations of: "model learning", "data-driven modeling", "sparse regression", "symbolic regression", "equation discovery", "system identification", and "SINDy". This was complemented by backward snowballing from key references and extending to publications citing them. Articles were selected based on three criteria: (i) ability to learn both structure and parameters of models of an ODE system, (ii) relevance to challenges related to biological modeling, (iii) originality of the method. For symbolic regression, only about 20\% met these criteria while the majority of the reviewed sparse regression methods met these criteria.

\paragraph{Shiny Application}
To interactively navigate, filter, and visualize the methods reviewed in this article, we developed a Shiny application (Shiny for Python). Its modular design enables easy updates, offering a dynamic complement to the static review for researchers seeking suitable modeling tools. Each entry is annotated with:
\begin{itemize}
\item \textbf{Bibliographic information:} based on Google Scholar.
\item \textbf{General method characteristics:} method name, method type (symbolic or sparse regression), concise description of the method's strategy, types of equations identified (ODEs, PDEs, etc.), and biological systems or benchmarks used for evaluation.
\item \textbf{Biological and methodological challenges:} handle missing data, handle unobserved species, incorporation of prior knowledge, handle of multiple experimental conditions or individuals, preprocessing procedures applied, and the types of candidate functions used.
\item \textbf{Technical aspects:} optimization algorithm, hyperparameter tuning strategy, use of Bayesian inference, use of deep learning architectures, and uncertainty quantification capabilities.
\item \textbf{Implementation details:} Programming language and URL to the public implementation (if available).
\end{itemize}

\section{Acknowledgment}
We sincerely thank Lucie Gaspard-Boulinc (Institut Curie, INSERM U1331) for her contribution in the development of the Shiny application and for her help with the visualization of the results of this review.

\bibliographystyle{natbib}
\bibliography{ref}

\end{document}